\documentclass[apj]{emulateapj}
\usepackage{amssymb,latexsym,amsmath}
\usepackage{apjfonts}
\usepackage{graphicx}
\usepackage{psfig}

\def \nh {N${\rm _H}$}
\def \ergsec{\hbox{erg s$^{-1}$}}

\def \arcmin {\hbox{$^\prime$}}
\def \arcsec {\hbox{$^{\prime\prime}$}}

\def\spose#1{\hbox to 0pt{#1\hss}}
\def\ltsim{$\mathrel{\spose{\lower 3pt\hbox{$\sim$}}
        \raise 2.0pt\hbox{$<$}}$\thinspace}
\def\gtsim{$\mathrel{\spose{\lower 3pt\hbox{$\sim$}}
        \raise 2.0pt\hbox{$>$}}$\thinspace}
\def \msun {${\rm M_\odot}$}

\def \nh {$N_{\rm H}$}

\def \dtwentyfive {${\rm D_{25}}$}

\newcommand\tvir{{\hbox{$T_{\rm vir}$}}}
\newcommand{\source}{\mbox{NGC\,5044}}

\newcommand{\apec}{APEC}
\newcommand{\mekal}{MEKAL}
\newcommand\emin{\hbox{{$E_{\rm min}$}}}

\newcommand{\zfe }{${\rm Z_{Fe}}$}

\newcommand{\chandra }{{\em Chandra}}
\newcommand{\spitzer }{{\em Spitzer}}
\newcommand{\xspec }{{\em Xspec}}

\newcommand{\ciao }{{\em CIAO}}

\newcommand{\heasoft }{{\em Heasoft}}

\newcommand{\xmm }{{\em XMM}}

\newcommand{\rosat }{{\em ROSAT}}

\newcommand{\ned}{{\em{NED}}}

\def \dtwentyfive {${\rm D_{25}}$}

\newcommand\omegam{\hbox{{$\Omega_{\rm m}$}}}
\newcommand\omegalambda{\hbox{{$\Omega_{\Lambda}$}}}
\newcommand\kmsmpc{{\rm km s$^{-1}$ Mpc$^{-1}$}}
\newcommand\ho{\hbox{{$H_{0}$}}}
\newcommand\thot{\hbox{{$T_{\rm h}$}}}
\newcommand\tcool{\hbox{{$T_{\rm c}$}}}
\slugcomment{Submitted to the Astrophysical Journal}
\shorttitle{The X-ray cavities, filaments and cold fronts of NGC 5044}
\shortauthors{Gastaldello et~al.}

\begin{document}

\title{The X-ray cavities, filaments and cold fronts in the core of the galaxy group\\ NGC 5044}

\author {Fabio Gastaldello\altaffilmark{1,2,3},
         David A. Buote\altaffilmark{3},
	 Pasquale Temi\altaffilmark{4,5},	
         Fabrizio Brighenti\altaffilmark{6,7}, 
         William G. Mathews\altaffilmark{7}
	 \& Stefano Ettori\altaffilmark{8,9}		
}
\altaffiltext{1}{IASF-Milano, INAF, via Bassini 15, Milano 20133, Italy}
\altaffiltext{2}{Occhialini Fellow}
\altaffiltext{3}{Department of Physics and Astronomy, University of
California at Irvine, 4129 Frederick Reines Hall, Irvine, CA 92697-4575}
\altaffiltext{4}{Astrophysics Branch, NASA/Ames Research Center, MS 245-6, Moffet Field, CA 94035}
\altaffiltext{5}{SETI Institute, Mountain View, CA 94043; and Department of physics and Astronomy, University of Western Ontario, London ON N6A, 3K7, Canada}
\altaffiltext{6}{Dipartimento di Astronomia, Universit\`a di Bologna, via
Ranzani 1, Bologna 40127, Italy}
\altaffiltext{7}{UCO/Lick Observatory, University of California at Santa Cruz,
 1156 High Street, Santa Cruz, CA 95064}
\altaffiltext{8}{INAF, Osservatorio Astronomico di Bologna, via
Ranzani 1, Bologna 40127, Italy}
\altaffiltext{9}{INFN, Sezione di Bologna, viale Berti Pichat 6/2, I-40127 Bologna, Italy}
\begin{abstract}
We present a two-dimensional analysis of the bright nearby galaxy group 
NGC 5044 using the currently available \chandra\ and \xmm\ data. In the 
inner 10 kpc a pair of cavities are evident together with a set of bright 
X-ray filaments. If the cavities are interpreted as gas 
displaced by relativistic 
plasma inflated by an AGN, even in the absence of extended 1.4 GHz 
emission, this would be consistent with a recent outburst as also indicated
by the extent of dust and H$\alpha$ emission. The soft X-ray filaments 
coincident with H$\alpha$ and dust emission are cooler than the ones which do 
not correlate with optical and infrared emission. We suggest that dust-aided
cooling contributes to form warm ($T \sim 10^4$ K) gas, emitting H$\alpha$ 
radiation.
At 31 kpc and 67 kpc a pair of cold fronts are present, indicative of 
sloshing due to a dynamical perturbation caused by accretion of a less 
massive group, also suggested by the peculiar velocity of the brightest 
galaxy NGC 5044 with respect to the mean group velocity.
\end{abstract}


\section{Introduction} 
\label{Introduction} 
%
The current X-ray observatories, \chandra\ and \xmm, have revolutionized
our understanding of the cores of relaxed galaxies, groups and clusters,
which show a highly peaked X-ray emission from a hot interstellar medium whose 
radiative cooling time is less than 1 Gyr \citep[for recent determinations of 
cooling times in these objects see for example][]{Voigt.ea:04,Sanderson.ea:06,Jetha.ea:07}.
In the absence of heating, a cooling flow is established, in which the gas  
cools, condenses and flows toward the center, accreting onto the central 
galaxy \citep{Fabian:94}. However the mass sink for all this supposedly 
cooling and condensing gas has never been 
found \citep{Donahue.ea:04}. X-ray observations with {\em Chandra} and 
{\em XMM} have established that there is little evidence for emission from 
gas cooling below $\sim$ \tvir/3: just when gas should be cooling most 
rapidly it appears not to be cooling at all 
\citep[see the recent review by][and references therein]{Peterson.ea:06}.

A compensating heat source must therefore resupply the radiative losses, and
many possibilities have been proposed, including thermal conduction 
\citep[e.g.,][]{Narayan.ea:01}, energy released by mergers 
\citep[e.g.,][]{Motl.ea:04} or by supernovae \citep[e.g.,][]{Silk.ea:86}. 
However, feedback from the central AGN
has become the most appealing solution to the problem 
\citep[see the recent review by][and references therein]{McNamara.ea:07}.
There is, in fact, clear observational evidence for AGN heating as 
the majority of brightest central galaxies of clusters and groups host a 
radio loud AGN \citep[e.g.,][]{Burns:90,Best.ea:07} and, following the launch 
of \chandra, in an increasing number of objects
such disturbances as shocks, ripples and cavities have been found in the
central atmospheres of clusters, groups and elliptical galaxies 
\citep[e.g.,][]{Fabian.ea:06,Birzan.ea:04,Dunn.ea:06,Forman.ea:05,Vrtilek.ea:02,Jetha.ea:07,Croston.ea:08,Finoguenov.ea:01*2,Allen.ea:06}. 
The cavities, which appear as X-ray
surface brightness depressions, have been interpreted as bubbles of
low density relativistic plasma inflated by radio jets, displacing the thermal
gas causing $PdV$ heating \citep[e.g.,][]{Churazov.ea:02}, although \citet{Mathews.ea:08}
pointed out that cavity formation contributes thermal energy that may
offset radiative cooling only by injecting ultra-hot but non relativistic gas;
X-ray cavities formed solely from relativistic gas have a global cooling effect.
In most cases, the energy introduced by
the AGN is more than sufficient to counteract putative cooling flows 
\citep{McNamara.ea:07}, although the physical process of the coupling of the 
feedback energy with the ambient medium is not well understood. 
While many of the observed X-ray cavities are filled 
with plasma emitting in the radio at 1.4 GHz, some are undetected at this 
frequency and have been referred to as ``ghost cavities''. 
These may result from 
the aging of the relativistic particle population and be the signature of a 
previous AGN outburst. Observations at low radio frequency of several rich 
clusters with ghost cavities show that they are indeed filled with 
relativistic plasma \citep[e.g., A 2597,][]{Clarke.ea:05}, although even
low frequency emission cannot be easily detected.

However, some net cooling is probably occurring, as the galaxies at the 
center of cool cores show properties  not shared by typical 
elliptical galaxies \citep{Crawford:04}:
1) the presence of strong,
low-ionization emission line nebulae both in  
clusters \citep[e.g.,][]{Crawford.ea:99,Conselice.ea:01} and in groups 
\citep[e.g.][]{Macchetto.ea:96}; 2) the galaxies at 
the center of clusters that display these nebular emission show a component 
of excess ultraviolet/blue continuum associated with young stars
\citep[e.g.,][]{Cardiel.ea:98,Crawford.ea:99}. This excess blue light is again
extended on scales of several kpc and there is evidence that some 
(spatially extended) star
formation has been triggered by interaction with the radio source
\citep{Mcnamara:04} 3) CO emission lines have been detected in several
cool core clusters at millimeter wavelengths 
\citep[e.g.,][]{Edge:01,Salome.ea:03},
implying the presence of a substantial amount of warm molecular gas
($10^{9-11.5}$ M$_\odot$) within a 50 kpc radius of the central
galaxy. Less massive warmer molecular regions have been observed in
H$_2$ lines, which are often spatially associated with the stellar UV
and H$\alpha$ emission extending over $\sim 20$ kpc \citep{Jaffe.ea:05}.

The origin and excitation mechanism of the nebular H$\alpha$ emission
have been much debated and they are still poorly known. The nebulae
require a constant and distributed heating source
\citep[e.g.][]{Johnstone.ea:88} which could be stellar \citep[e.g. massive OB
stars,][]{Allen.ea:92} or the ICM \citep[e.g. conduction,][]{Sparks.ea:89}. 
Deep \chandra\ X-ray imaging of Perseus \citep{Fabian.ea:03*1} and M87 
\citep{Sparks.ea:04} have renewed interest in conduction because of the 
spatial coincidence of filamentary H$\alpha$ and soft X-ray emission, which 
can be due to conduction and mixing of the cold gas with the ICM. 
The disposition of some of
the optical filaments in Perseus strongly suggests that the filaments
are due to the buoyant radio bubbles drawing out the cold gas from a
central reservoir  \citep{Fabian.ea:03*1} and it reveals the possibility of
tracing the flow of the rising bubbles \citep{Hatch.ea:06}.  But the
question is still open if galaxy-galaxy interactions are required to
stimulate the central optical nebulae and if the optical filaments are
trails of galaxies punching through a molecular hydrogen reservoir as
suggested by the examples in \citet{Wilman.ea:06}. The excitement caused by
the shift in the cooling flow paradigm is stimulating also a renewed
theoretical interest in the quest for the origin and excitation of
molecular \citep{Ferland.ea:08} and H$\alpha$ emission
\citep{Nipoti.ea:04,Pope.ea:08,Revaz.ea:08}. 
In particular \citet{Nipoti.ea:04} have
shown how the H$\alpha$ nebulae can be stable against thermal
evaporation only for the conditions of temperature and pressure found
in cool cores, explaining their association with that type of
environment.

High resolution \chandra\ observations have also revealed another interesting
and unexpected feature in the ICM of relaxed clusters: the presence of
cold fronts. Many clusters have been found to exhibit sharp arc-shaped
jumps in their gas density and temperature which, unlike shock fronts,
have the gas on the dense side cooler, so the pressure is continuous across
the front \citep[e.g.,][]{Markevitch.ea:00,Vikhlinin.ea:01}. In merging 
systems they were interpreted as contact discontinuities between gases from
different sub-clusters \citep{Markevitch.ea:00}. However cold fronts are present
in the centers of many, if not most, relaxed clusters with cool cores
\citep[e.g.,][]{Mazzotta.ea:01,Markevitch.ea:03*1,Dupke.ea:03,Ascasibar.ea:06,Ghizzardi.ea:07}. 
The widely accepted scenario is that cold fronts in these systems are due to
sloshing of the cool gas in the central gravitational potential, which is
set off by minor mergers/accretions; the only necessary condition is a steep
entropy profile as observed in relaxed clusters \citep{Ascasibar.ea:06}.

Whereas a growing number of clusters and elliptical galaxies have deep enough 
multi-wavelength data (X-rays, radio and optical) to study the rich 
phenomenology of cool cores in a detailed spatially resolved fashion, only an 
handful of groups with such coverage exists, as for example the GEMS objects 
\citep{Osmond.ea:04} presented in \citet{Rasmussen.ea:07}, and therefore,
``unfortunately, AGN heating is not as well studied in groups as in clusters''
\citep{McNamara.ea:07}. Examination of AGN feedback at the mass scale of groups
is valuable because, although the scale of outbursts in groups is less
energetic and often on a smaller spatial scale than in clusters, the impact can
be even more dramatic than in rich clusters due to
the shallower group potential. Statistical studies examining the impact of AGN
on groups are starting to address the points raised above 
\citep{Croston.ea:05,Jetha.ea:07}.
Cold fronts have not been investigated in detail so far in relaxed groups of galaxies.

In this paper we show with more detail the currently available 
\chandra\ and \xmm\
data for the galaxy group NGC 5044, one of the brightest groups in X-rays:
it is one of the only 5 objects with ${\rm{k}}T < 2$ keV in the HIFLUGCS sample 
\citep{Reiprich.ea:02}. Our estimate, using the new \chandra\ and \xmm\ data, 
for the bolometric (0.1-100 keV) X-ray luminosity within
$r_{500}=443 h_{70}^{-1}$ kpc \citep{Gastaldello.ea:07*1} is 
$1.05\pm0.06 \times 10^{43} h_{70}^{-2}$ \ergsec, in good agreement with the \rosat\ estimate 
presented in \citet{Reiprich.ea:02} of $2.46 \times 10^{43} h_{50}^{-2}$ within 
$560 h_{50}^{-1}$ kpc.  
In our previous analysis, mainly focused on 
radial, azimuthally averaged properties of the system \citep{Buote.ea:03,Buote.ea:03*1}, 
interesting features like holes and filaments within 10 kpc, a sharp edge, 
resembling a cold front, at 67 kpc, and unusually low iron abundances at 
large radii \citep{Buote.ea:04} were already pointed out.
An H$\alpha$ nebula is present in the core of NGC 5044 showing an extended
filamentary structure \citep{Goudfrooij.ea:94*1,Caon.ea:00}.
NGC 5044 is also remarkable because {\em Spitzer} data show extended cold dust
emitting at 70 $\mu m$ \citep{Temi.ea:07}. Moreover, \citet{Temi.ea:07*1} showed
extended 8 $\mu m$ excess (likely arising from PAH, polycyclic aromatic hydrocarbon, molecules) 
extending out to several kpc and
spatially coincident with the H$\alpha$ emitting nebulosity and the 
brightest soft X-ray emission. 
As proposed in \citet{Temi.ea:07*1}, current evidence 
is consistent with an internal origin of this dust, which has been buoyantly 
transported from the galactic core out to several kpc into the hot X-ray emitting gas 
following an AGN outburst.

All distance-dependent quantities have been computed assuming \ho = 70 \kmsmpc,
\omegam = 0.3 and \omegalambda = 0.7. At the redshift of $z=0.009$ 1\arcmin\
corresponds to 11.1 kpc. All the errors quoted are at the 68\% confidence
limit.
%
\section{Observations and data preparation}
\label{obs}
%
\begin{figure}[th]
\centerline{\psfig{figure=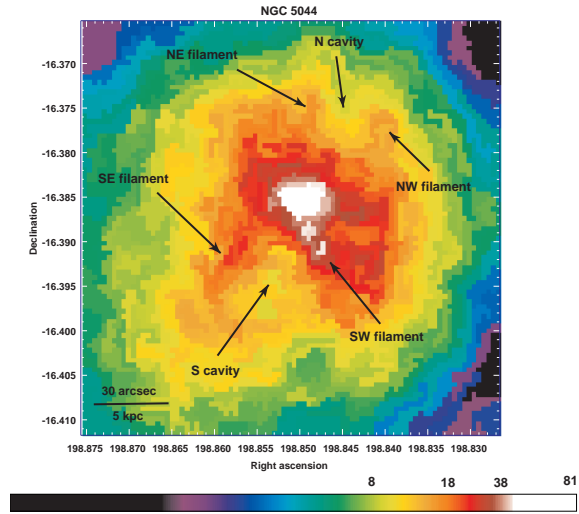,height=0.3\textheight}}
\caption{\label{fig.source} \footnotesize \chandra\ 0.5-5 keV X-ray image of 
the inner 30x30 kpc of \source. The image has been processed to remove point 
sources, flat fielded with a 1.0 keV exposure map. The figure has 
been contour binned using the algorithm of \citet{Sanders:06} with a 
S/N of 10. Color bar units are total counts per pixel. Interesting features discussed
in the text are highlighted.
}
\end{figure}
%
%
\source\ has been observed by \chandra\ with the ACIS-S configuration on 
September 1999 (obsID 798) for 20.7 ks and by \xmm\ on January 2001 (obsID
0037950101) with the EPIC MOS and pn cameras for 23 and 20 ks, respectively
\citep[see also][]{Buote.ea:03,Buote.ea:03*1}. 
We focused mainly on the \chandra\ data because
its PSF is much better suited for the high-spatial resolution study that
constitutes most of the paper. We take advantage of the \xmm\ data when we
discuss the properties at larger radii (see section \S\ref{subsection_edges} and 
\S\ref{subsection_spectral.coldfronts}) 
where the resolution requirements are not so important and we can benefit from the
larger accessible field.
Here we provide just a brief description of the data preparation, more details 
can be found in \citet{Gastaldello.ea:07*1}.

\subsection{Chandra} 
The data were analyzed with the X-ray analysis packages
\ciao\ 3.4 and \heasoft\ 6.4 in conjunction with the \chandra\ calibration 
database (\emph{Caldb}) version 3.4.2. In order to ensure the most 
up-to-date calibration, all data were reprocessed from the ``level 1'' events 
files, following the standard \chandra\ data-reduction
threads\footnote{{http://cxc.harvard.edu/ciao/threads/index.html}}.
We applied the standard corrections to take account of a time-dependent 
drift in the detector gain and charge transfer inefficiency, as implemented 
in the \ciao\ tools.  From low surface brightness regions of the active chips
we extracted a light-curve (5.0-10.0~keV) to identify and excise 
periods of enhanced background. The observation was quiescent resulting in a 
final exposure time of 20 ks. 
Point source detection was performed using the \ciao\
tool {\tt wavdetect} and removed, so as not to contaminate the diffuse
emission, using appropriate elliptical regions containing 99\% of their flux.  
We generated an image in the 0.5-5.0 keV and a corresponding exposure map 
computed at an energy of 1.0 keV. A zoom of the image in the inner region 
is presented in Fig.\ref{fig.source} and Fig.\ref{fig.cavities} and the the region 
covered by the ACIS-S3 chip is presented in Fig.\ref{fig.innercoldfrontsb}.

\subsection{XMM}
We generated calibrated event files with SAS v7.1.0 using the tasks
{\em emchain} and {\em epchain}. We considered only event patterns 0-12 for 
MOS and 0 for pn, and the data were cleaned using the standard procedures for 
bright pixels and hot column removal and pn out-of-time correction.
Periods of high background due to soft protons were filtered as in 
\citet{Gastaldello.ea:07*1} resulting in a net exposure time of 22 ks for 
MOS1 and MOS2 and 17ks for pn. For each detector, we created images in the 
0.5-2 keV band with point sources detected using the task {\tt ewavelet} and 
masked using circular regions of 25  radius centered at the source position. 
We created exposure maps for each detector and we combined the MOS images 
into a single exposure-corrected image, smoothed on a scale of 10\arcsec,
shown in Fig.\ref{fig.xmmcoldfront}. 
\section{X-ray images and surface brightness profiles}
\label{images}

\subsection{The inner 10 kpc}\label{subsection.core}
\begin{figure}[t]
\centerline{\psfig{figure=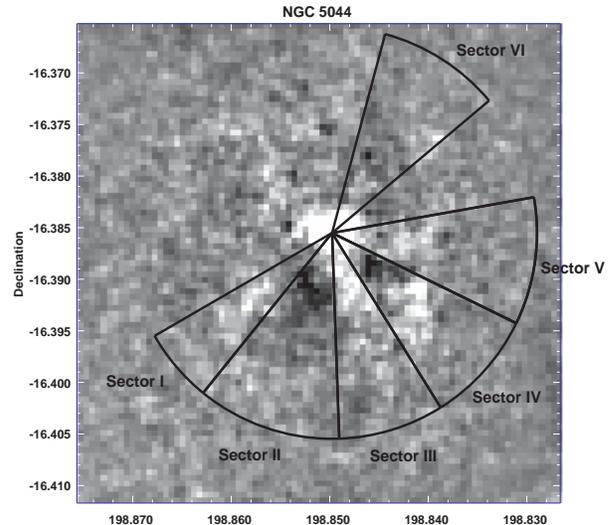,height=0.3\textheight}}
\caption{\label{fig.umask} \footnotesize Unsharp-masked \chandra\ image created by
subtracting a 0.5-5 keV intensity image smoothed by a $\sigma = 10$\arcsec\
Gaussian and one smoothed by a $\sigma = 1$\arcsec\ Gaussian. Areas in black
show a deficit of counts, whereas areas in white show an excess of counts.
Also shown are the angular sectors used for the surface brightness profiles
shown in Fig.\ref{fig.profiles}, labeled by name according to the scheme of 
Table \ref{tab.profiles}.
}
\end{figure}

The presence of a disturbed morphology 
with filamentary structure is already evident from the raw \chandra\ image and 
further confirmed by using the contour binning 
technique of \citet{Sanders:06} (See Fig.\ref{fig.source}).
Two depressions in surface brightness and multiple filamentary structures, some
of them connected to the presence of the cavities, have been highlighted in 
Fig.\ref{fig.source}. To highlight the presence of structure we performed an
``unsharp masking'' of the \chandra\ image \citep[e.g.,][]{Fabian.ea:03}: the 0.5-5 keV 
exposure-map-corrected image was smoothed with Gaussians of width 1 and 10 
arcsec and the two smoothed images were then subtracted, with the resulting 
image shown in Fig.\ref{fig.umask}. The southern cavity and the structure of filaments 
are clearly shown in the unsharp masked image. To quantify the observed structures we 
extracted from the exposure corrected image (the unsharp image was
not used in the following analysis of the surface brightness profiles) 
an azimuthally averaged radial profile and profiles from selected angular sectors
as depicted in Fig.\ref{fig.umask} and listed in Table \ref{tab.profiles}  
\citep[centered on the X-ray surface brightness peak coincident with the 
optical center of the galaxy NGC 5044 as listed by \ned\ and by][RA 13:15:24.0 and DEC -16 23 09, J2000 coordinates]{Macchetto.ea:96}.

%
%
\begin{deluxetable}{ll}
\tablecaption{\label{tab.profiles}}
\tabletypesize{\scriptsize}
\tablehead{\colhead{Sector} & \colhead{PA}}
\startdata
Sector I & 120-141 \\
Sector II & 141-182 \\
Sector III & 182-212 \\
Sector IV & 212-244 \\
Sector V & 244-280 \\
Sector VI & 310-345 \\
\enddata
\tablecomments{Position angles (measured form the N direction) for the surface brightness profiles discussed in the text in \S\ref{subsection.core}.}
\end{deluxetable}
%
\begin{figure*}[th]
\centerline{
\parbox{0.5\textwidth}{
\psfig{figure=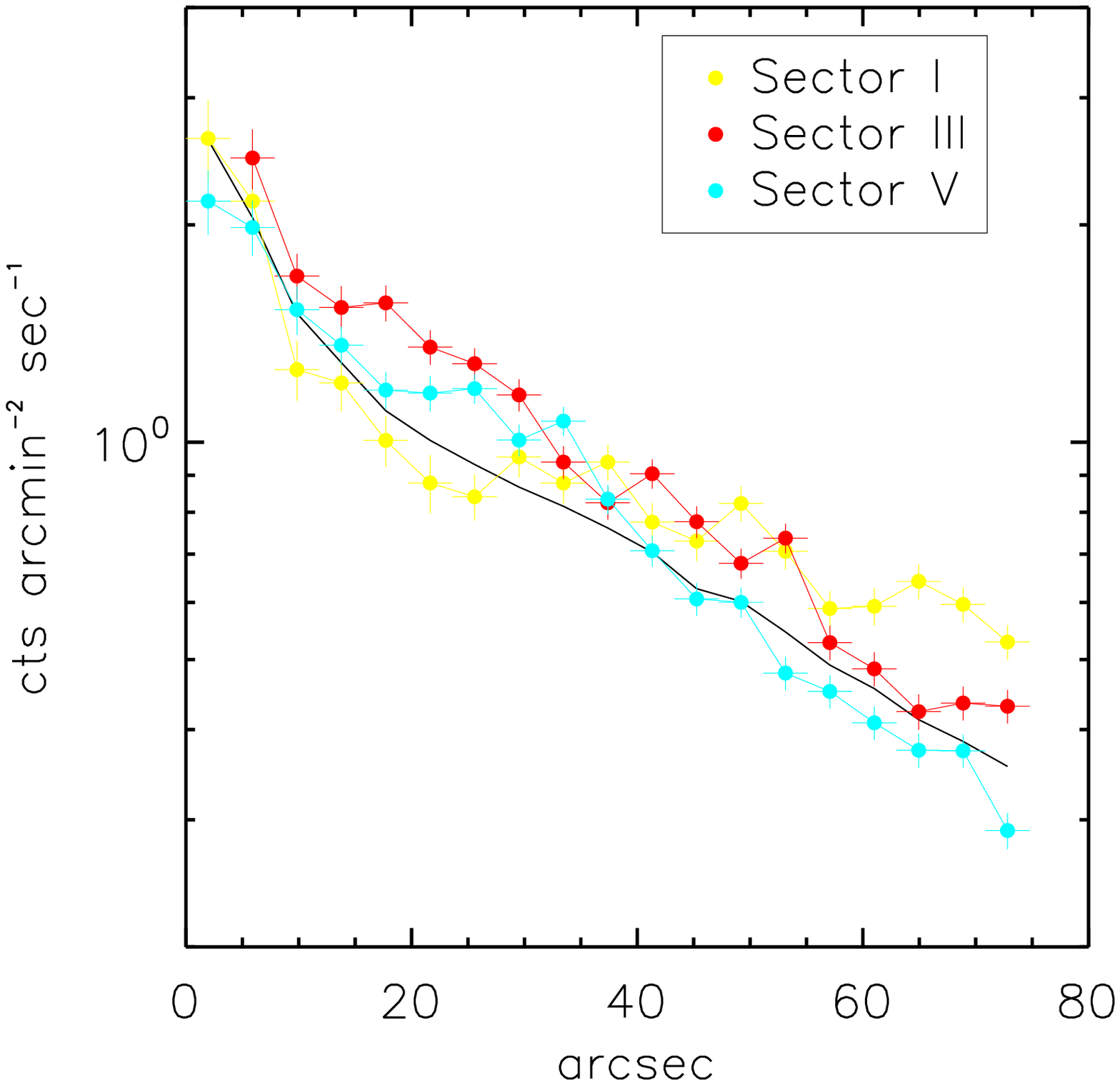,height=0.28\textheight}}
\parbox{0.5\textwidth}{
\psfig{figure=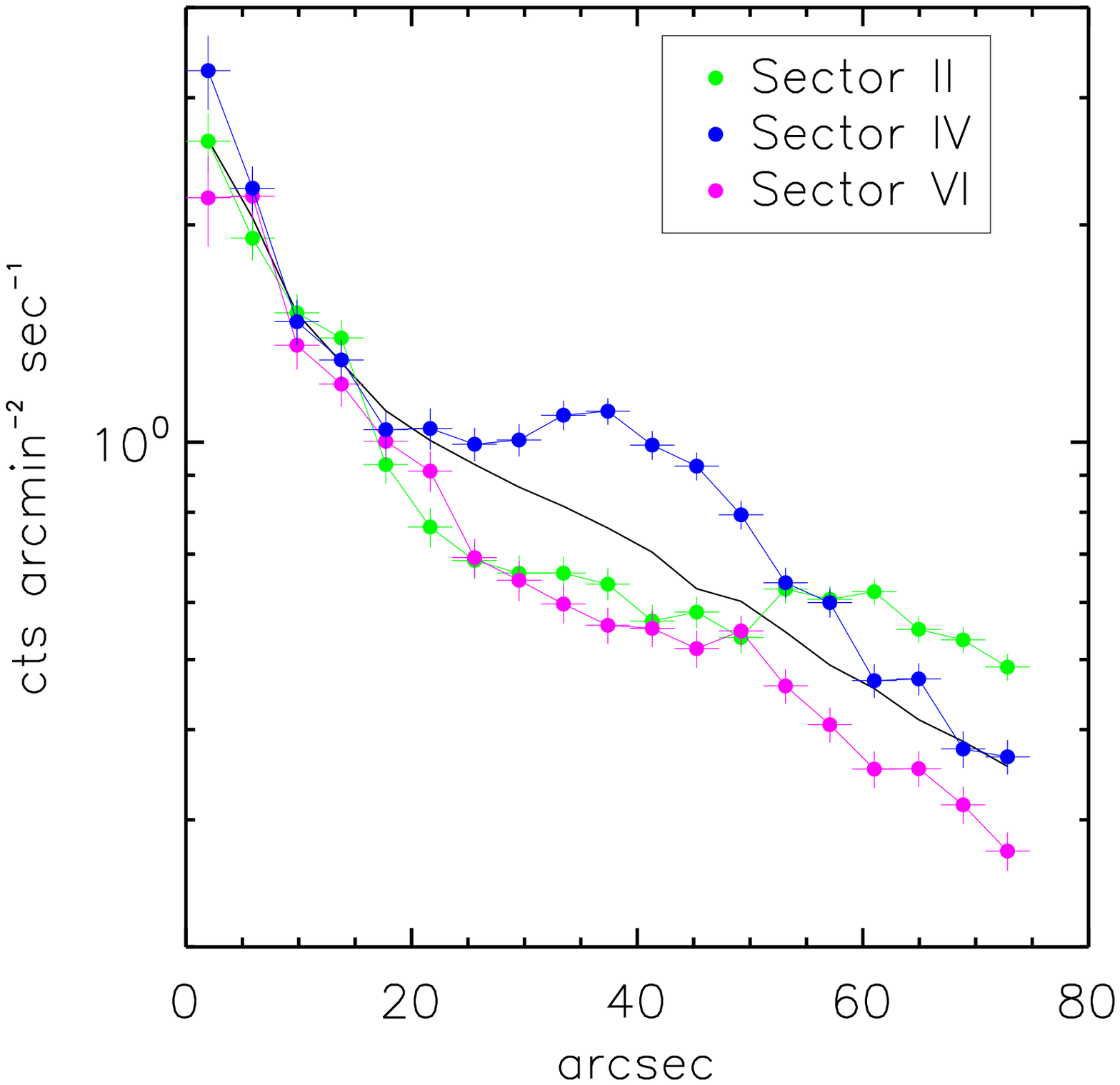,height=0.28\textheight}}
}
\centerline{
\parbox{0.5\textwidth}{
\psfig{figure=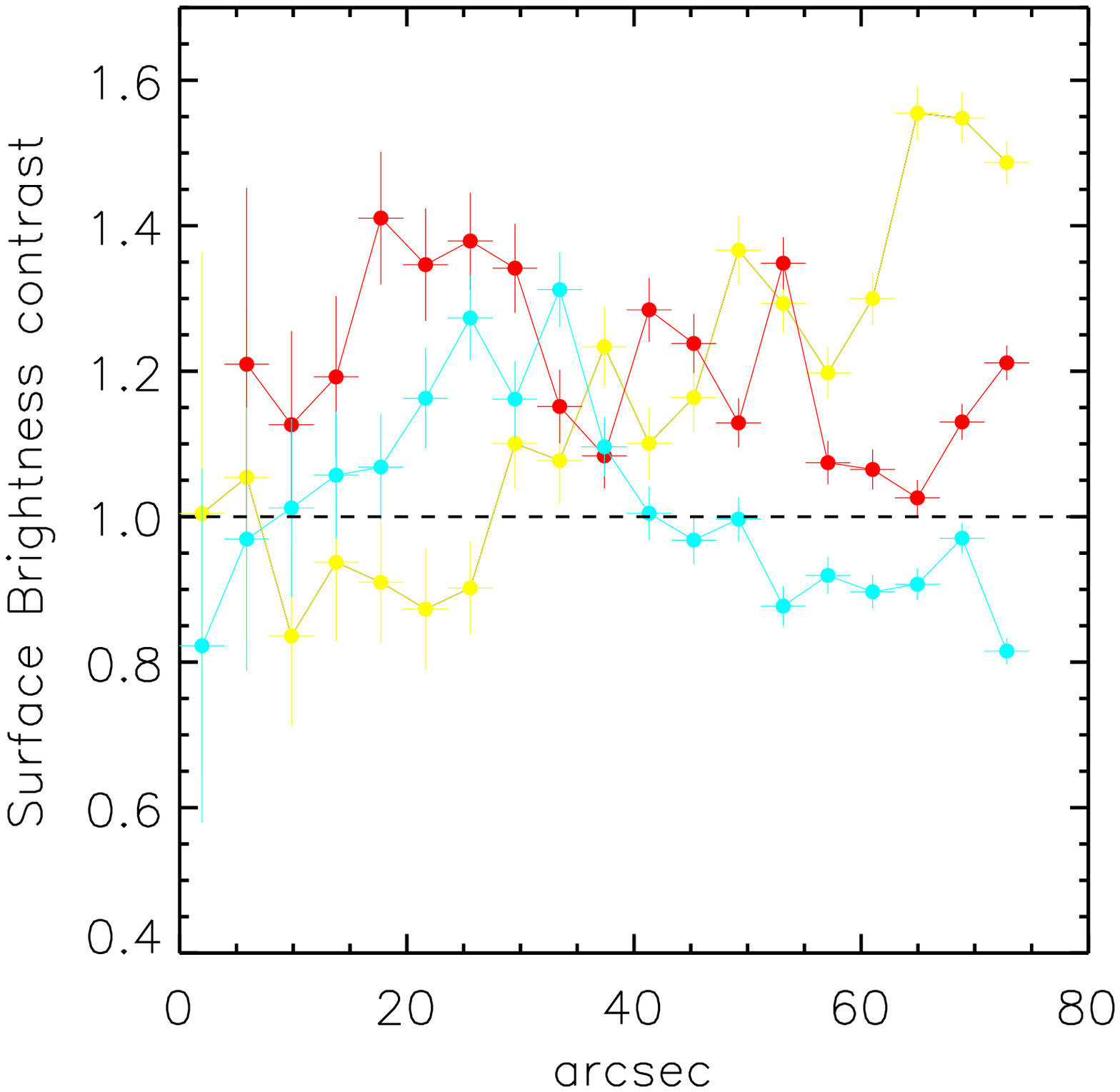,height=0.28\textheight}}
\parbox{0.5\textwidth}{
\psfig{figure=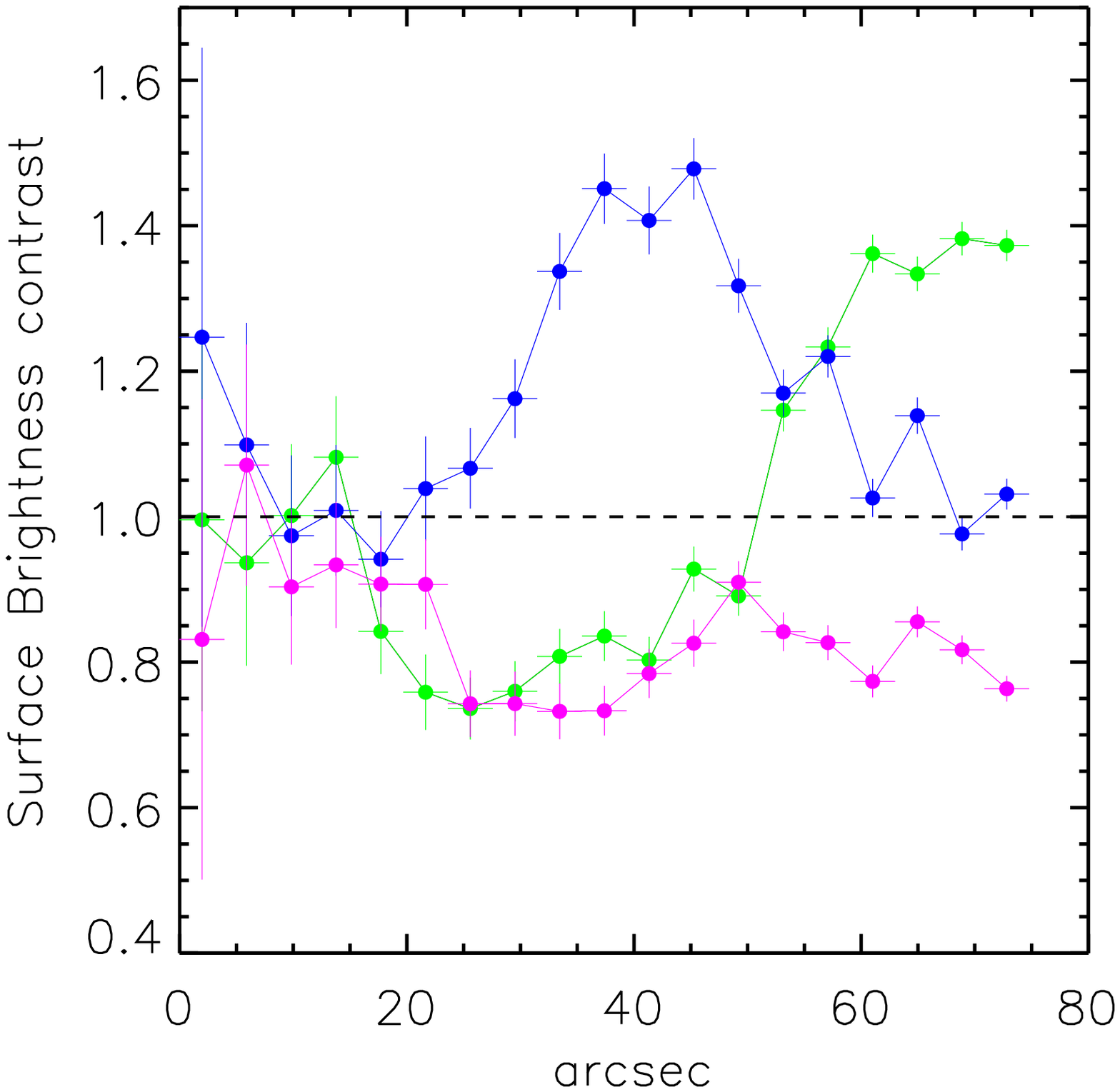,height=0.28\textheight}}
}
\caption{\label{fig.profiles} \footnotesize
\emph{Top panel:} Surface brightness profiles in selected angular sectors listed in
Table \ref{tab.profiles} and shown in Fig.\ref{fig.umask}. The azimuthally averaged 
surface brightness profile is 
plotted as a solid black line in both panels for ease of comparison.\\
\emph{Bottom panel:} Ratio of the surface brightness profiles of the above 
selected angular sectors over the azimuthally averaged one.
}
\end{figure*}

As it can be seen from Fig.\ref{fig.profiles}, there is a complex network of
filamentary structure of enhanced emission which highlights dramatically 
regions of comparatively lower surface brightness. Two filaments, brighter and
extended (in particular compared to the azimuthally averaged profile, also 
shown in the right panel of Fig.\ref{fig.profiles}), are evident in sector I
(the SE filament) and sector III (the SW filament) with an embedded cavity in 
Sector II (also in this sector the emission is more extended compared to the average, 
i.e. there is an excess of counts for radii greater than 1\arcmin). The filament in 
sector III is brighter in the region 10\arcsec-30\arcsec\ but less extended, 
i.e. dimmer at $r>1$\arcmin, than  the overall emission at PA 120-182. 
The emission in sector IV can robustly be considered filamentary, with 
enhanced emission, even brighter than the other two filaments in the range 
30\arcsec-50\arcsec\ but less extended, with an embedded depression in surface brightness 
between 10\arcsec and 30\arcsec, which stands out as a region of low emission, compared to
the surroundings, in Fig.\ref{fig.umask}. However this feature is at the same level of 
surface brightness of the azimuthally averaged profile.
The cavity in sector VI is the real counter-part of the one in sector II. 
Although it stands out less clearly in the unsharp-mask image, it is 
clearly indicated in the binned images of Fig.\ref{fig.source} and in the 
striking similarity of the surface brightness profiles within 50\arcsec\ as 
shown in Fig.\ref{fig.profiles}. The northern cavity stands out less clearly
in the unsharp masked image of Fig.\ref{fig.umask} because of the dimmer regions 
surrounding it (the northern arms are not as bright as the southern ones) 
and because of the sharper drop in surface brightness at $r>50$\arcsec. 
There is also evidence of enhanced emission in sector V in the 
20\arcsec-40\arcsec\ region, as highlighted also in the unsharp masked image.

\subsection{The inner and outer surface brightness edges}\label{subsection_edges}

\begin{figure*}[th]
\centerline{
\parbox{0.5\textwidth}{
\psfig{figure=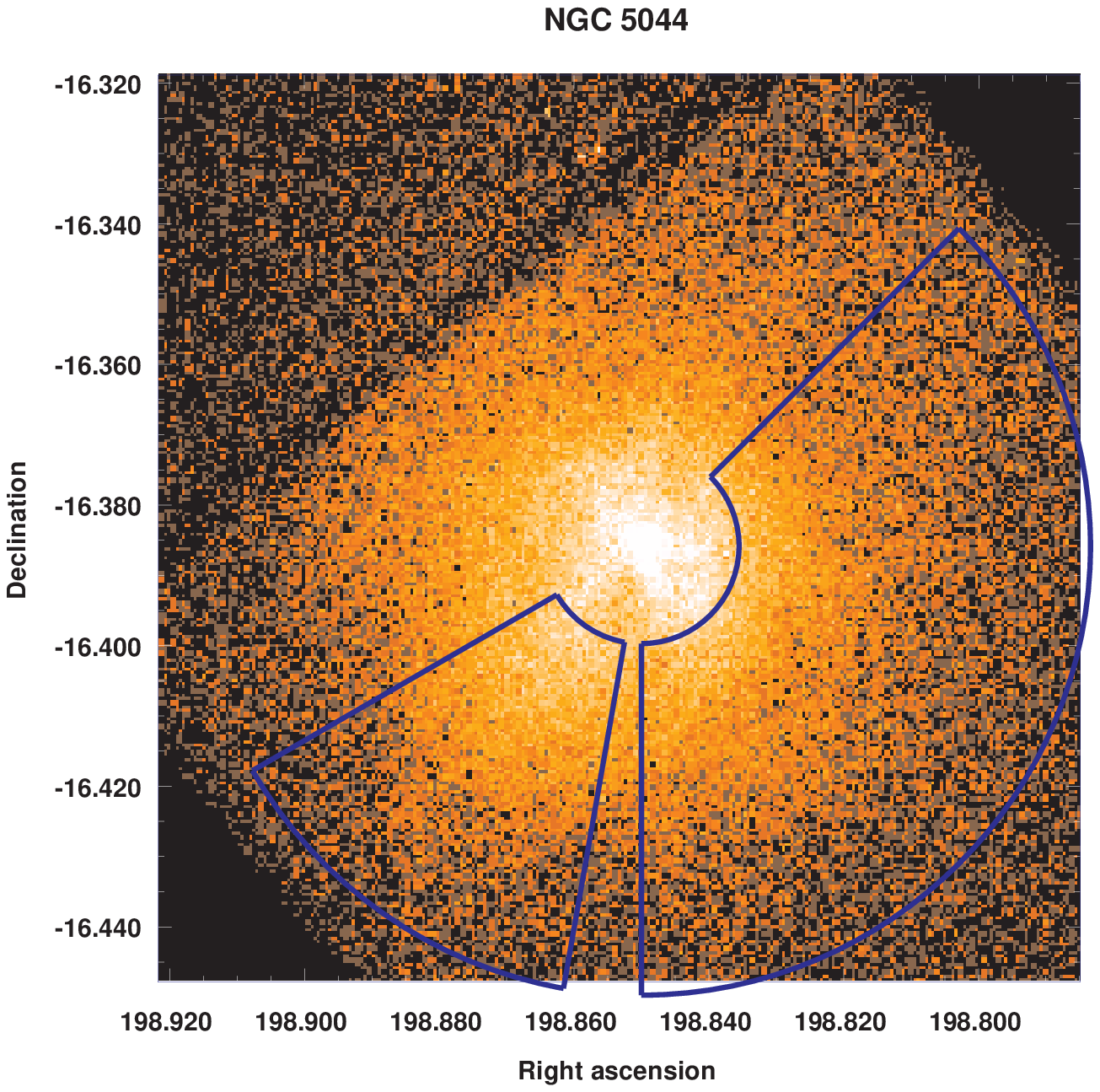,height=0.28\textheight}}
\parbox{0.5\textwidth}{
\psfig{figure=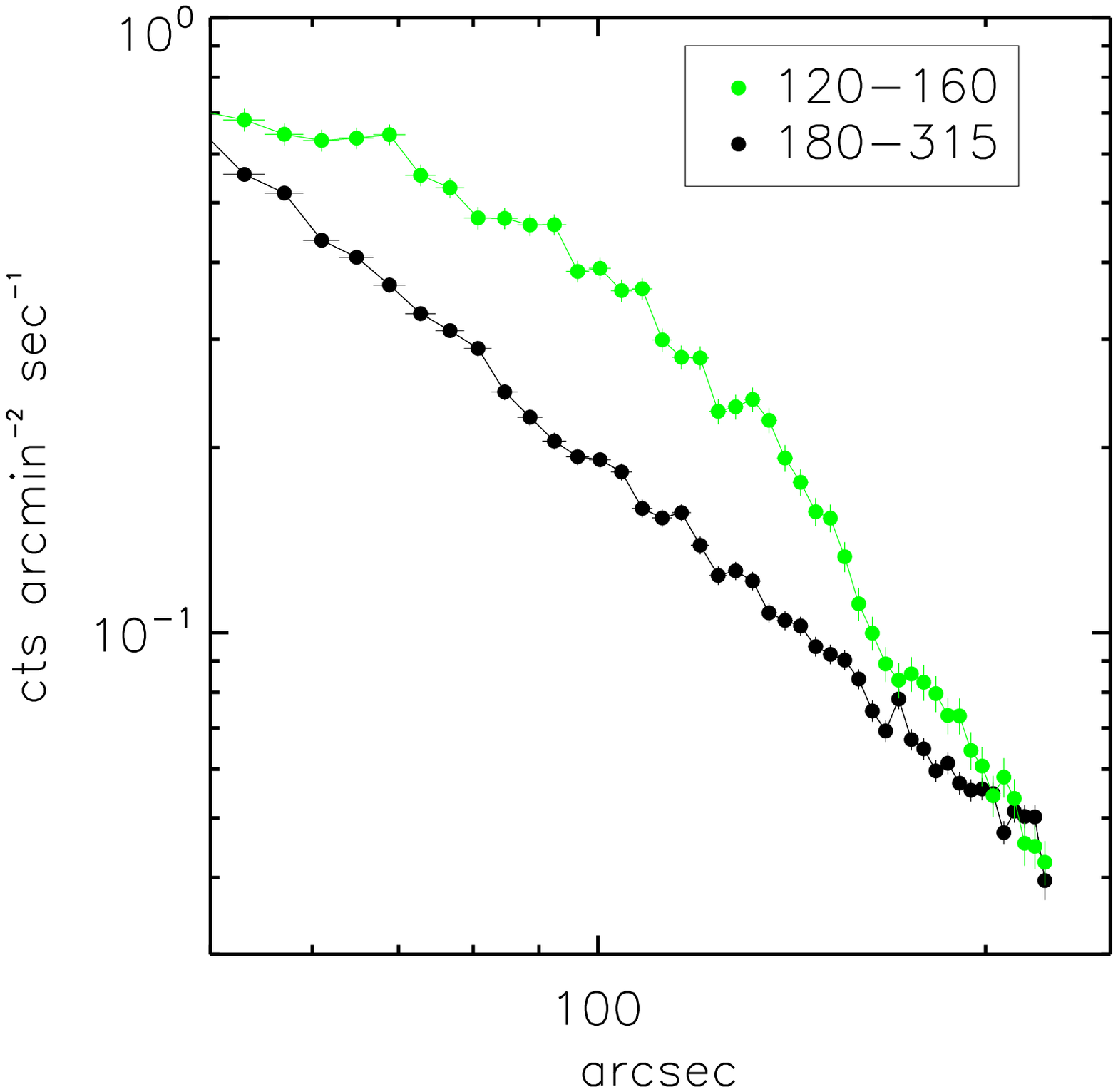,height=0.28\textheight}}
}
\caption{\label{fig.innercoldfrontsb} \footnotesize
\emph{Left panel:} Over-plotted on the \chandra\ image are the sector used
to extract the surface brightness profile across the cold front and the
reference sector used for comparison.\\
\emph{Right panel:} Surface brightness profile for the cold front and the
reference sector. A sharp drop is visible at $\sim$ 170\arcsec\ form the 
center, i.e. 31 kpc.
}
\end{figure*}

We now zoom out to investigate the outer regions in the \chandra\
image, as shown in Fig.\ref{fig.innercoldfrontsb}. A sharp edge in the surface
brightness on the south-east (PA 120-160) is revealed in this image and
further confirmed by a surface brightness profile in the same sector, as 
compared to a reference profile extracted in the PA 180-315 sector (right
panel of Fig.\ref{fig.innercoldfrontsb}). Whereas the profile is smooth
and with a continuous derivative in the reference sector, the slope of the
sector across the cold front is rapidly changing and the edge is 
at $\sim 31$ kpc (170\arcsec).

\begin{figure}[t]
\centerline{\psfig{figure=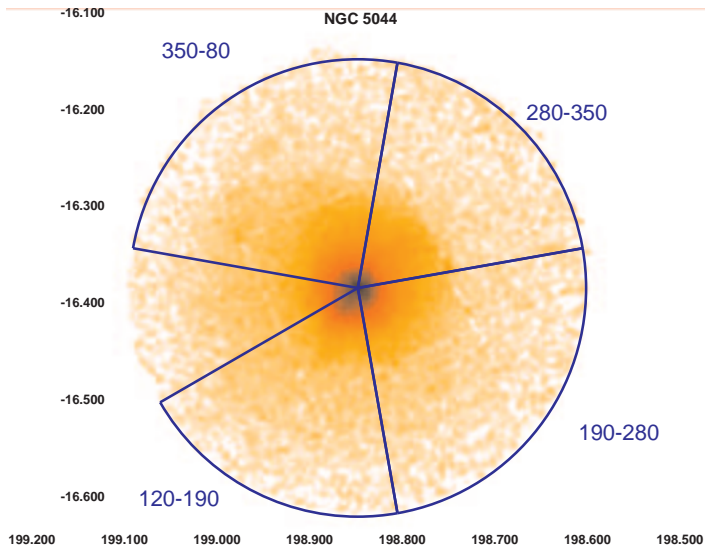,height=0.3\textheight}}
\caption{\label{fig.xmmcoldfront} \footnotesize Mosaic of the MOS1 and MOS2 images 
smoothed on a 10\arcsec\ scale. The image has been divided by the summed 
exposure maps to correct for exposure variations. Point sources have been
removed and filled using the \ciao\ task {\tt dmfilth}. Sectors used for the
extraction of surface brightness profiles discussed in the text and presented
in Fig.\ref{fig.xmmprofiles} are also shown.
}
\end{figure}

\begin{figure*}[th]
\centerline{
\parbox{0.5\textwidth}{
\psfig{figure=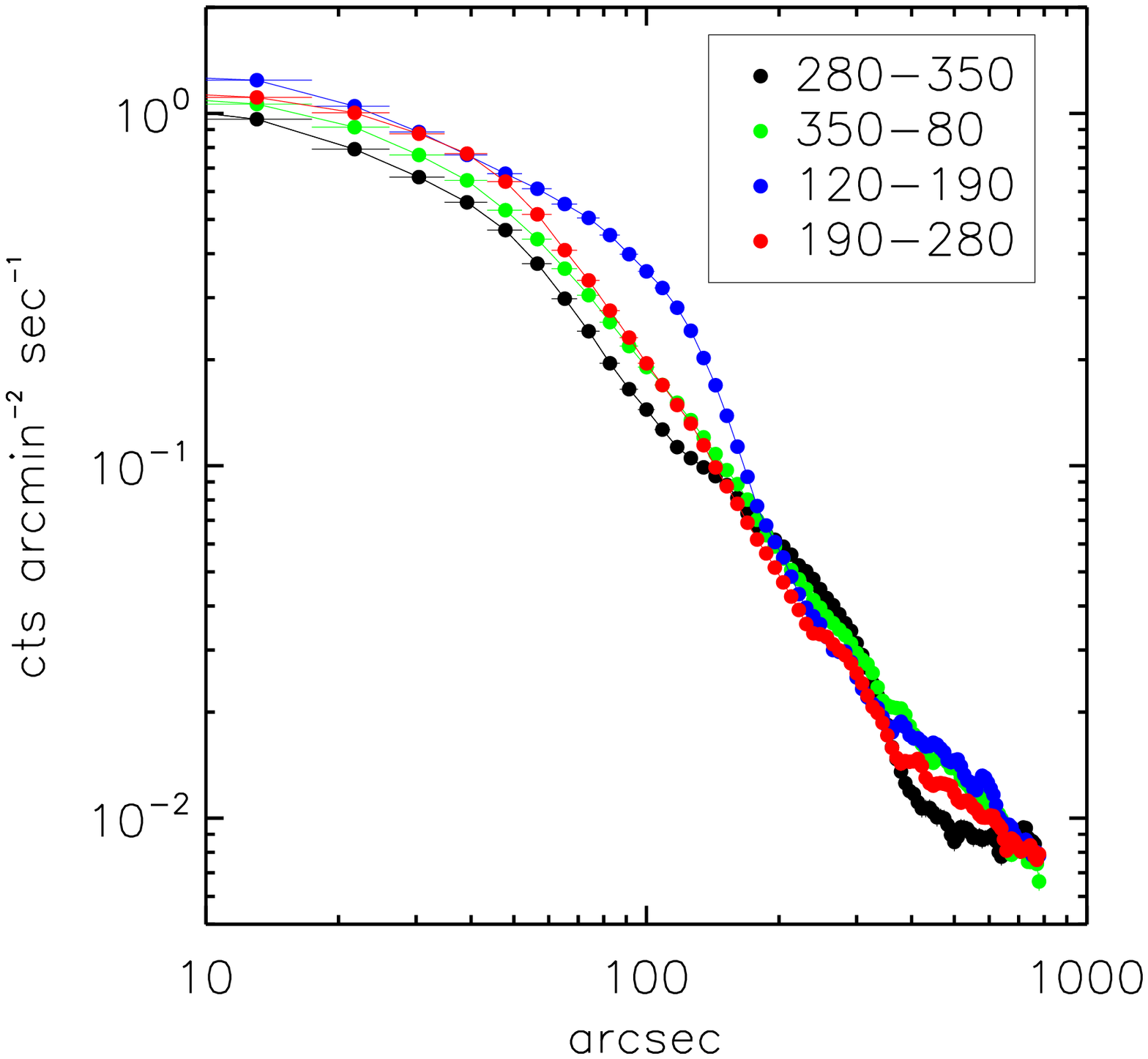,height=0.28\textheight}}
\parbox{0.5\textwidth}{
\psfig{figure=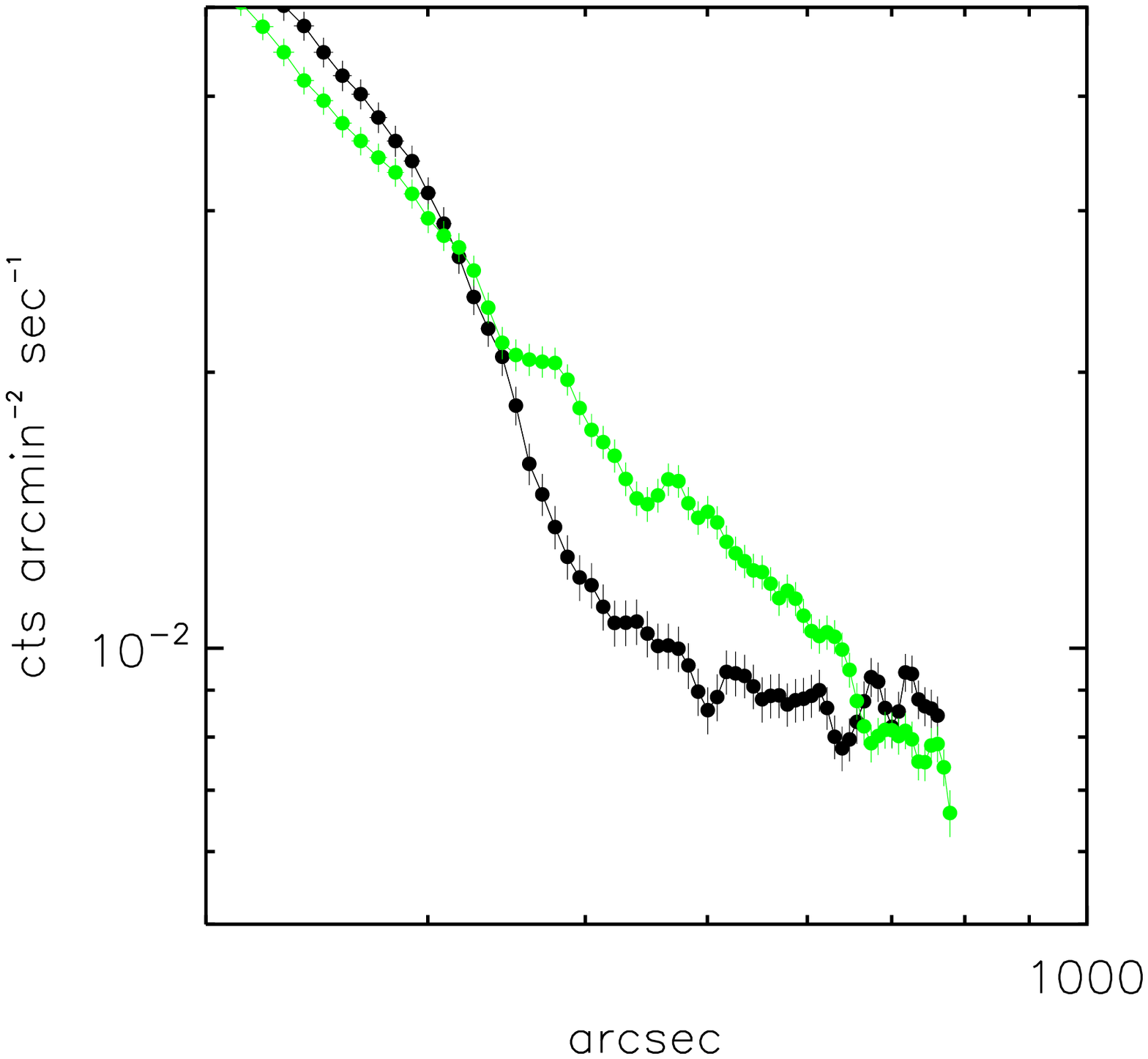,height=0.28\textheight}}
}
\caption{\label{fig.xmmprofiles} \footnotesize
\emph{Left panel:} Surface brightness profiles in selected angular sectors, 
with PA measured from the N direction. The inner edge can be easily seen, 
though not as sharp as in the \chandra\ profile because of the \xmm\ larger
PSF.\\
\emph{Right panel:} Zoom over the interesting radial region for the outer
edge. The characteristic shape of a rapid change in the slope of the profile
across the edge can easily be seen in the sector PA 280-350. The profile
in the undisturbed sector PA 350-80 is plotted for comparison.
}
\end{figure*}

Inspection of the combined MOS image shown in Fig.\ref{fig.xmmcoldfront} 
confirms the suggestion of \citet{Buote.ea:03*1} of the presence of another 
surface brightness edge in the north-west direction at $\sim$ 6\arcmin. Plots 
of the surface brightness profiles across this sector and in other three
sectors (one including the inner edge) for comparison are shown in
Fig.\ref{fig.xmmprofiles}. The inner cold front is detected also in the MOS
image. More importantly another sharp feature (considering also the
\xmm\ PSF at this off-axis angle, $\sim 50$\arcsec\ 90\% encircled energy
fraction at 1.5 keV) is clearly revealed at $\sim$350\arcsec, i.e. 65 kpc 
(see right panel of Fig.\ref{fig.xmmprofiles}). 
The feature is detected also in the pn data.

%
%

\section{Spectral analysis and temperature maps}

Details about the \chandra\ and \xmm\ spectral extraction and analysis
can be found in \citet{Humphrey.ea:06} and \citet{Gastaldello.ea:07*1}.
Here we briefly summarize that for \chandra\ appropriate count-weighted 
spectral response matrices were generated for each region using the 
standard \ciao\ tasks {\tt mkwarf} and {\tt mkacisrmf} and for \xmm\
using the SAS tasks {\tt rmfgen} and {\tt arfgen} in extended source mode.
For each spectrum, we estimated the background through local modeling 
using the method outlined in \citet{Humphrey.ea:06} and 
\citet{Gastaldello.ea:07*1}. The spectra were re-binned to ensure a S/N of at 
least 3 and a minimum 20 counts bin$^{-1}$.

\subsection{The inner 10 kpc}\label{subsection_spectral.core}

\begin{figure*}[th]
\centerline{
\parbox{0.5\textwidth}{
\psfig{figure=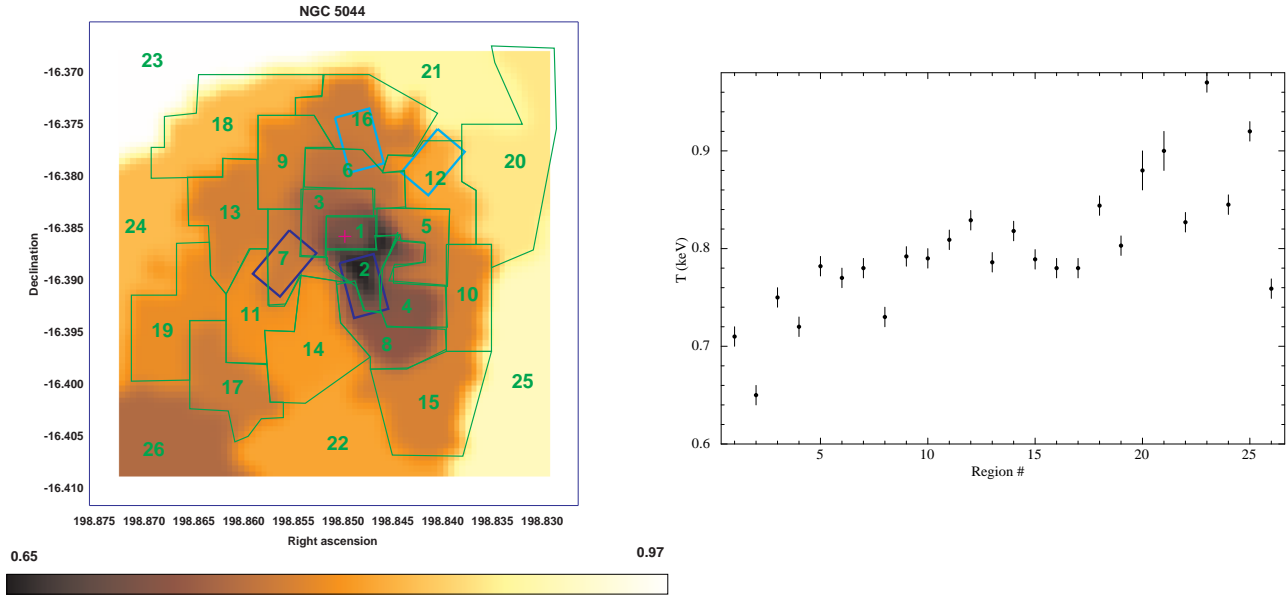,height=0.35\textheight}}
\parbox{0.5\textwidth}{
\psfig{figure=f7b.ps,angle=-90,width=0.45\textwidth}}
}
\caption{\label{fig.tmap} \footnotesize
\emph{Left panel:} Temperature map of the core regions of \source\ based on 
the SN=50 contour binning. The map has been smoothed for better presentation 
with a Gaussian kernel of 6\arcsec. Color bar units are in keV. The magenta cross marks 
the center of NGC 5044. The spectra 
presented in Fig.\ref{fig.comparison} and discussed in the text have been extracted from the 
rectangular regions shown in blue and they encompass the SE and SW filament; the spectra
extracted from the cyan boxes and discussed in the text encompass the NE and NW filaments
highlighted in Fig.\ref{fig.source};.  
\emph{Right panel:} Values and $1\sigma$ error bars for the temperature in 
each region of the temperature map.
}
\end{figure*}

We extracted spectra from the \chandra\ dataset in a series of regions as depicted in 
Fig.\ref{fig.tmap}, determined using the contour binning technique \citep{Sanders:06}
 with a S/N=50. 
We fitted the background subtracted spectra with an APEC thermal plasma model
\citep{Smith.ea:01} with the absorbing column density fixed at the Galactic 
value \citep{Dickey.ea:90}. Solar abundances are in the units of \citet{Grevesse.ea:98}.
To account for the undetected point-sources we 
added a 7.3 keV bremsstrahlung component 
\citep[this model gives a good fit to the spectrum of the detected sources in 
nearby galaxies,][]{Irwin.ea:03} for all the regions, because they fall
within the twenty-fifth magnitude isophote (\dtwentyfive) of the central 
galaxy NGC 5044.

The temperature map thus obtained is shown in Fig.\ref{fig.tmap}. The clearest
feature is the cool gas present in the SW (Sector III) arm (cooler than any 
other region in the map) which is spatially coincident with the H$\alpha$ and 
dust filament \citep[][see Fig.\ref{fig.halpha}]{Temi.ea:07*1}. 
The SE arm (Sector I) is hotter. 
There is presence of cooler gas with respect to the surroundings in the
direction of the inner surface brightness edge (i.e. region 26 in 
Fig.\ref{fig.tmap}). 

One possible concern about the interpretation of the spectral fits and
consequently of the surface brightness features is the suggestion of
limited multiphase gas in the temperature structure of \source, as derived
by the preference of two temperature (2T) models over single temperature (1T)
models in radial annuli \citep{Buote.ea:03}. This is further complicated by 
projection effects, given the fact that we are dealing only with projected
spectra in the analysis presented here: a proper deprojection is problematic 
due to uncertainty in the projection geometry and it is beyond the scope of the present paper.
However it is evident in the radial analysis
of \citet{Buote.ea:03} that the cool component clearly dominates in the inner
10 kpc (see their Figure 6) and that a single phase description, modulo
projection effects, is an appropriate description of the data for these 
inner regions. We investigated
fitting 2T models as in \citet{Buote.ea:03} with the addition of the
bremsstrahlung component for the unresolved point sources. The results
of the two-dimensional analysis are consistent with the radial analysis of
\citep{Buote.ea:03} and with the dominance in terms of emission measure of the
cool component over the hot one.

\begin{figure}[t]
\centerline{\psfig{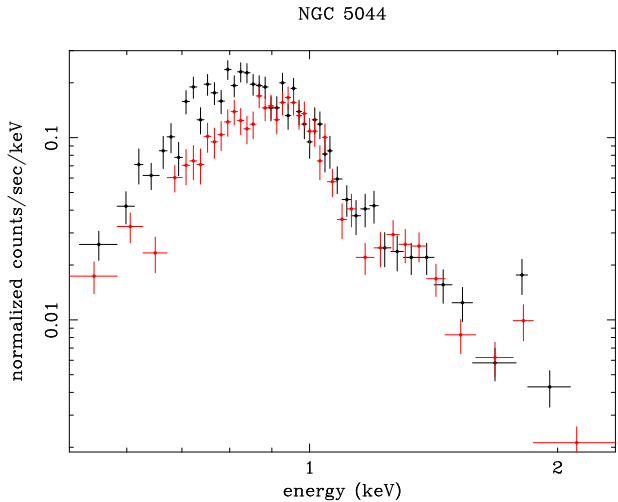}}
\caption{\label{fig.comparison} \footnotesize The spectra from the rectangular 
boxes taken respectively from the SW arm (black data points) and the SE arm
(red data points). Note the shape of the Fe-L ``hump'' and the prominence of 
lower energy lines, signature of lower temperature, in the spectrum of the SW
arm.
}
\end{figure}

To investigate in more detail the nature of the bright soft X-ray filaments
we extracted spectra representative of the two southern 
filaments in two 30\arcsec$\times$45\arcsec\ rectangular regions at a distance of 
30\arcsec\ from the center (the blue boxes shown in Fig.\ref{fig.tmap}); the spectrum 
taken from the SW filament has a Fe-L feature with lower
excitation energy lines more prominent than the spectrum from the SE filament, 
signature of a lower temperature (see Fig.\ref{fig.comparison}). If
we fit the two spectra with a 1T model we obtain a temperature difference
which is significant at $9\sigma$ (kT = $0.65\pm0.01$ keV for the SE arm 
and $0.78\pm0.01$ keV for the SW arm). The regions connected to the filaments are cooler 
compared to all the other regions in the map (by $3.6\sigma$ compared to the central region \#1, 
which has the second lowest temperature in the map, $0.70\pm0.01$). 
If we try to fit with a 2T model
the spectrum of the SE arm we obtain only a small decrease in $\chi^2$
($\chi^2$/dof = 57/40 for the 2T model compared to 60/42 of the 1T model).
If we repeat the exercise for the regions of brighter emission surrounding
the northern cavity (the cyan boxes shown in Fig.\ref{fig.tmap} encompassing the 
northern filaments highlighted in Fig.\ref{fig.source}) we obtain $0.74\pm0.02$ keV for 
the NW filament and  $0.82\pm0.02$ keV for the NE filament.

%
%
\subsection{The surface brightness edges}\label{subsection_spectral.coldfronts}

\begin{figure*}[th]
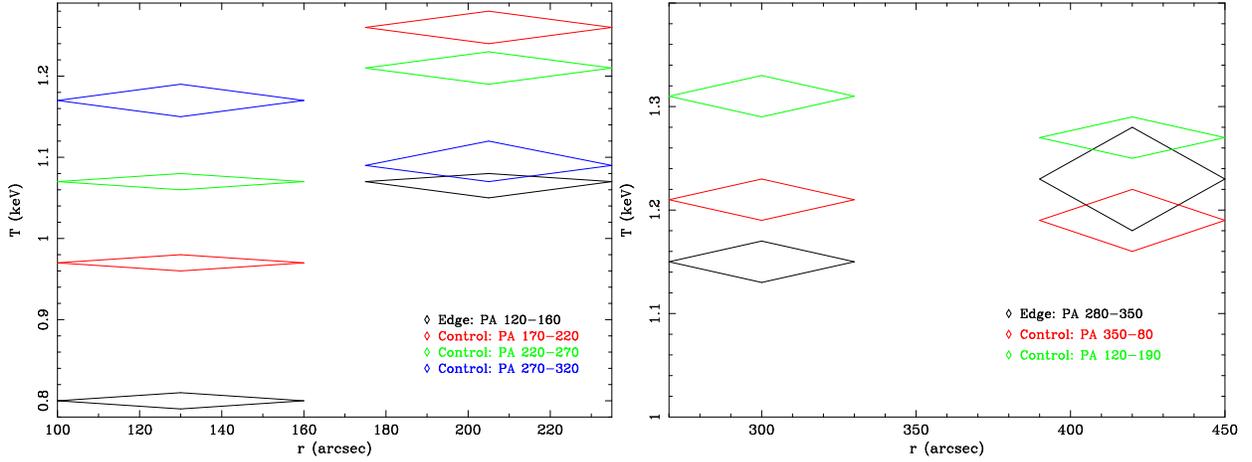

\centerline{
\parbox{0.5\textwidth}{
\psfig{figure=f9a.ps,angle=-90,height=0.25\textheight}}
\parbox{0.5\textwidth}{
\psfig{figure=f9b.ps,angle=-90,height=0.25\textheight}}
}
\caption{\label{fig.tjumps} \footnotesize
\emph{Left panel:} Values and $1\sigma$ error bars for the temperature across
the inner surface brightness edge and in control sectors.\\
\emph{Right panel:} Values and $1\sigma$ error bars for the temperature across
the outer surface brightness edge and in control sectors.
}
\end{figure*}

To investigate the nature of the surface brightness edges we have to
determine the temperature across the surface brightness jump. We therefore
extracted \chandra\ spectra for the inner jump together with 
``control' spectra taken from regions at the same radial distance from 
the center but at different position angles. We applied the same method to 
the outer edge when extracting \xmm\ spectra.
In Fig.\ref{fig.tjumps} we plot the temperature across the edges obtained by 
fitting a 1T model. In Table \ref{tab.coldfronts} we present the results
of 1T and 2T fits for selected regions inside and outside the edges together
with control regions at the same radial range but at different position 
angles. 

For the inner edge a clear temperature jump is detected, with the brighter
regions inside the edge cooler than the outer part. 2T models are preferred
over 1T models which points to some degree of projection effects in the
temperature determination across the edge, probably caused by the cool gas
which is producing the edge. Other regions covering the same range of radii but
at different position angles do not show in-fact the same behavior, with the
exception to some degree of the region immediately close to the edge. We extracted
\xmm\ MOS and pn spectra from the same regions inside and outside the inner edge finding
consistent results within the 1$\sigma$ errors for both 1T and 2T models.

For the outer edge the temperature jump is more subtle and of low statistical
significance ($1.6\sigma$), but, contrasted
with the slightly declining trend of the control regions
(we are in the region around the peak of the azimuthally averaged temperature 
profile), it is consistent with the picture of cooler gas inside the edge.

%
\begin{table*}[t] \footnotesize
\caption{Parameters from the spectral fits across the edges
\label{tab.coldfronts}}
\begin{center} \vskip -0.4cm
\begin{tabular}{ccccccccccc}
\tableline\tableline\\[-7pt]
& \multicolumn{4}{c}{1T} & \multicolumn{6}{c}{2T}\\
& & $T$ & \zfe\ & norm & & \tcool & \thot & \zfe\ & norm$_c$ & norm$_h$ \\
Radii PA & $\chi^2$/dof & (keV) & solar & ($10^{-4}$ cm$^{-5}$) & $\chi^2$/dof & (keV) & (keV) & solar & ($10^{-4}$ cm$^{-5}$) & ($10^{-4}$ cm$^{-5}$) \\
\tableline \\[-7pt]
{\emph{Inner edge}} \\
100-160\arcsec\  120-160 &   137/91  & $0.80 \pm 0.01$ & $0.59\pm0.07$ & $9.83^{+0.73}_{-0.97}$ & 88/89  & $ 0.77 \pm 0.01$ & $1.51 \pm0.16 $ & $1.20^{+0.28}_{-0.20}$ & $4.43^{+0.85}_{-0.79}$ & $1.81\pm0.25$ \\
175-235\arcsec\  120-160 &   87/66  & $1.06 \pm 0.01$ & $0.38\pm0.03$ & $5.64\pm0.31$ & 73/64  & $0.68^{+0.11}_{-0.07}$ & $1.24^{+0.07}_{-0.05}$ & $0.64^{+0.15}_{-0.10}$ & $0.62^{+0.43}_{-0.12}$ & $3.83^{+0.44}_{-0.65}$ \\
{\emph{Control regions}} \\
100-160\arcsec\  170-220 &   139/87  & $0.97 \pm 0.01$ & $0.77^{+0.13}_{-0.08}$ & $5.99^{+0.54}_{-0.73}$ & 118/85  & $ 0.85\pm0.01$ & $1.48^{+0.13}_{-0.61}$ & $1.52^{+0.48}_{-0.31}$ & $2.12^{+0.51}_{-0.48}$ & $1.78\pm0.30$ \\
175-235\arcsec\  170-220 &   70/62  & $1.27 \pm 0.02$ & $0.69^{+0.09}_{-0.07}$ & $3.96^{+0.23}_{-0.33}$ & \nodata  & \nodata & \nodata & \nodata & \nodata & \nodata \\
100-160\arcsec\  220-270 &   90/81  & $1.07 \pm 0.01$ & $0.86^{+0.13}_{-0.10}$ & $4.71\pm0.54$ & 80/79  & $ 1.06\pm0.01$ & \nodata\ & $1.52^{+0.64}_{-0.40}$ & $2.59^{+0.94}_{-0.77}$ & $0.88^{+0.22}_{-0.28}$ \\
175-235\arcsec\  220-270 &   69/66  & $1.21 \pm 0.02$ & $0.83^{+0.15}_{-0.10}$ & $3.73\pm0.46$ & \nodata  & \nodata & \nodata & \nodata & \nodata & \nodata \\
100-160\arcsec\  270-320 &   77/78  & $1.14 \pm 0.02$ & $0.88^{+0.17}_{-0.12}$ & $4.23^{+0.52}_{-0.55}$ & \nodata  & \nodata & \nodata\ & \nodata & \nodata & \nodata \\
175-235\arcsec\  220-270 &   73/73  & $1.09 \pm 0.02$ & $0.61^{+0.12}_{-0.08}$ & $4.70\pm0.52$ & \nodata  & \nodata & \nodata & \nodata & \nodata & \nodata \\
{\emph{Outer edge}} \\
270-330\arcsec\  280-350 &   214/178  & $1.15\pm 0.02$ & $0.40^{+0.07}_{-0.05}$ & $5.46^{+0.23}_{-0.26}$ & 207/176 & $ 0.66^{+0.16}_{-0.19}$ & $1.22^{+0.06}_{-0.03}$ & $0.51^{+0.28}_{-0.20}$ & $0.30^{+0.38}_{-0.13}$ & $4.71^{+0.32}_{-0.54}$ \\ 
390-450\arcsec\  280-350 &   83/73  & $1.23 \pm 0.05$ & $0.28^{+0.08}_{-0.05}$ & $2.59\pm0.33$ & \nodata  & \nodata & \nodata & \nodata & \nodata & \nodata \\
{\emph{Control regions}} \\
270-330\arcsec\  350-80 &   268/235  & $1.21\pm 0.02$ & $0.58^{+0.06}_{-0.05}$ & $5.98\pm0.37$ & \nodata  & \nodata & \nodata & \nodata & \nodata & \nodata \\
390-450\arcsec\  350-80 &   164/181  & $1.19\pm 0.03$ & $0.36^{+0.05}_{-0.04}$ & $5.55\pm0.41$ & \nodata  & \nodata & \nodata & \nodata & \nodata & \nodata \\
270-330\arcsec\  120-190 &   186/171  & $1.31\pm 0.02$ & $0.48^{+0.05}_{-0.04}$ & $4.58^{+0.20}_{-0.30}$ & 182/169 & $ 1.06^{+0.17}_{-0.12}$ & $1.68^{+0.14}_{-0.19}$ & $0.74^{+0.19}_{-0.16}$ & $1.11^{+1.45}_{-0.58}$ & $2.73^{+0.75}_{-1.11}$ \\ 
390-450\arcsec\  120-190 &   126/125  & $1.27\pm 0.02$ & $0.61^{+0.12}_{-0.09}$ & $3.07\pm0.32$ & \nodata  & \nodata & \nodata & \nodata & \nodata & \nodata \\
\tableline 
\end{tabular}
\tablecomments{Results of 1T and 2T spectral fits for selected regions as described in the
text in \S \ref{subsection_spectral.coldfronts}. The first and second columns refer
to the radial range (in arcseconds) and azimuthal range (PA in degrees) of the fitted 
sectors discussed in \S\ref{subsection_spectral.coldfronts}. The $norm$ parameter is the
emission measure of the \apec\ model as defined in \xspec:
$10^{-14}(\int n_en_pdV)/4\pi D^2(1+z)^2$ with units $\rm cm^{-5}$. 
\tcool and \thot refer to the temperature of the cool and hot component of the 2T model,
whereas norm$_c$ and norm$_h$ refer to the correspondent emission measures.
No entry for 2T models
means that there was no improvement over a 1T model. Regions with no entry for \thot\ 
did not significantly constrain that parameter.}
\end{center}
\end{table*}
%
%
\section{Systematic Errors}
\label{systematics}
This section contains an investigation of possible systematic errors
in the data analysis relevant for this paper. A thorough analysis of 
many of the issues involved has been conducted in \citet{Buote.ea:03} and 
\citet{Buote.ea:03*1}.

\subsection{Image Binning}
The presence of the morphological structures discussed in the paper,
in particular the disturbed morphology with filamentary structure
of the inner regions, is already evident from the raw \chandra\ image. 
The use of the particular binning technique of \citet{Sanders:06} does not
introduce any spurious feature. We also used the Weighted Voronoi Tessellation (WVT)
binning algorithm of \citet{Diehl.ea:06}, which is more
robust against the introduction of spurious features than adaptive smoothing,
as provided for example by the \ciao\ task {\tt csmooth} and we found the same features.

\subsection{Plasma Codes}
\label{plasma}

We compared the results obtained using the \apec\ code to those
obtained using the \mekal\ code \citep{Kaastra.ea:93,Liedahl.ea:95} to assess the
importance of different implementations of the atomic physics and
different emission line lists in the plasma codes.
We found no qualitative differences between the two codes: the fitted 
temperatures agree to within 5\% and abundances within 10-20\% (where 1T and 2T
models are compared accordingly) and well within their statistical errors.

\subsection{Bandwidth}

We explored the sensitivity of our results to our default lower limit of
the bandpass, $\emin=0.5$~keV. For comparison we performed spectral fits with 
$\emin=0.4$~keV and $\emin=0.7$~keV. The fitted temperatures are consistent
between models; iron abundances are better constrained, in particular for
2T models, when using the larger bandwidth \citep[see][]{Buote.ea:03*1}.

\subsection{Variable \nh}
\label{nh}
We take into account possible deviations for \nh\ from the value of 
\citet{Dickey.ea:90} allowing the parameter to vary by $\pm25$\%.
We found no qualitative differences between the two cases.
If we leave \nh\ free to fit we derive higher values in the core, but
still consistent at 1$\sigma$ with the Galactic value in the majority of the 
regions of Fig.\ref{fig.tmap}. All the other spectral parameters are basically
unchanged. This is not due to excess absorption but just the exploitation of
an additional free parameter by the fitting program to model deviations from
a single temperature model.  

\subsection{Background}

Since NGC 5044 is sufficiently bright and the temperature determination stems 
from the measurement of the shape and width of the Fe-L shell which is a 
prominent feature in the spectra, the fitted temperature values are quite 
insensitive to errors in the background normalization.

%
\section{Discussion}
\label{discussion}
\subsection{The nature of the cavities and filaments in the inner 10 kpc}
%
\begin{figure}[t]
\centerline{\psfig{figure=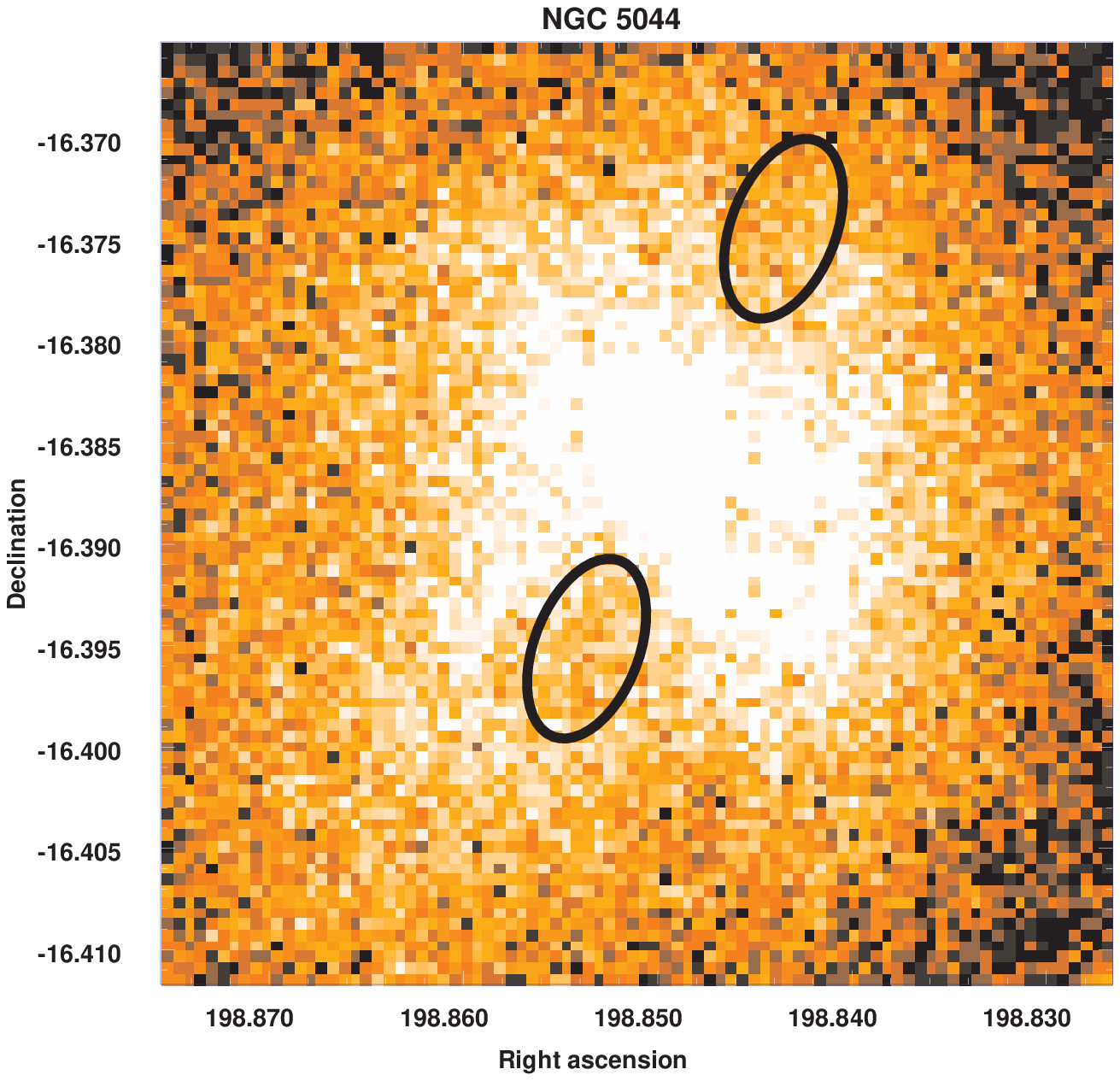,height=0.3\textheight}}
\caption{\label{fig.cavities} \footnotesize Exposure corrected 0.5-5 keV image of 
\source. Elliptical regions in black represent the shape and size of the 
detected X-ray cavities.
}
\end{figure}
%
%
\def\arraystretch{1.1}
\begin{deluxetable*}{cccccccc}
\tabletypesize{\scriptsize}
\tablewidth{0pt}
\tablecaption{Cavities time scales and derived powers. \label{tab.powers}}
\tablehead{
\colhead{$pV$\tablenotemark{a}} & \colhead{$t_{sound}$} & \colhead{$t_{buoy}$\tablenotemark{b}} & \colhead{$t_{ref}$\tablenotemark{b}}  & \colhead{$P_{sound}$\tablenotemark{c}} & \colhead{$P_{buoy}$\tablenotemark{c}} & \colhead{$P_{ref}$\tablenotemark{c}} & \colhead{$L_{X}$\tablenotemark{d}}\\
\colhead{($10^{55}$ erg)} & \colhead{($10^7$ yr)} & \colhead{($10^{7}$ yr)} &  \colhead{($10^7$ yr)} & \colhead{($10^{41}$ \ergsec)} & \colhead{($10^{41}$ \ergsec)} &  
\colhead{($10^{41}$ \ergsec)} & \colhead{($10^{41}$ \ergsec)}
}
\startdata
 $4.58$ & $1.34$  & $1.20$  & $3.50$ & $1.08$ & $1.21$ & $0.41$ & $32.1\pm0.09$ \\
        &         & $0.69$  & $2.02$ &        & $2.10$ & $0.72$ &                \\
\enddata
\tablenotetext{a}{It refers to the sum of the $pV$ energies of the two cavities}
\tablenotetext{b}{The first value refer to $g$ calculated
according to the hydrostatic equilibrium mass estimate whereas
the second entry refer to the estimate made using the stellar velocity dispersion of the
central galaxy, as described in the text.} 
\tablenotetext{c}{The power is the $pV/t$ work only, with $\gamma/(\gamma-1)$ not accounted for.
For a fully relativistic plasma the values for the powers need to be multiplied by 4.} 
\tablenotetext{d}{Bolometric (0.1-100 keV) X-ray luminosity within a radius of 27 kpc, where
the cooling time is 3 Gyr.}
\end{deluxetable*}

The X-ray analysis shown in this paper reinforces the scenario proposed 
for \source\ by \citet{Temi.ea:07*1}, which we briefly summarize below.
In a recent survey of elliptical galaxies observed with the {\it Spitzer} telescope 
\citet{Temi.ea:07} found spatially extended cold interstellar dust emitting at
70 $\mu$m around many group-centered X-ray luminous elliptical galaxies.
The source of this dust is the dust-rich cores of the central galaxy, not from mergers
\citep{Mathews.ea:03,Temi.ea:07}. Since this dust is thought to be in direct contact with
the hot, virialized interstellar gas ($kT \sim 1$ keV), it has a short lifetime ($\sim 10^7$ yrs)
to sputtering destruction by thermal ions.
Consequently, this dust is a spatial tracer of extremely transient events that recently 
occurred on kpc scales. \source, together with NGC 4636, has been investigated in more detail
in \citet{Temi.ea:07*1}. \source\ is not only extended at 70 $\mu$m but also 
at 8 $\mu$m in a manner similar to the
highly asymmetric H$\alpha$ optical line emission from warm gas.
Current evidence is consistent with the hypothesis that dust has been buoyantly transported 
from the galactic cores out to several kpc following a feedback heating
event. Furthermore, disorganized fragments of optically absorbing dusty gas
are visible in {\it HST} images within the central $\sim$100 parsecs,
providing further evidence of a recent central energy release. In particular, 
the association of interstellar PAH emission
(responsible of the 8 $\mu$m emission) and 
warm gas ($T \sim 10^4$ K) in \source\ indicates that we may be
viewing this galaxy at a rare moment immediately following
a release of energy near the central black hole.

\begin{figure*}[th]
\centerline{
\parbox{0.5\textwidth}{
\psfig{figure=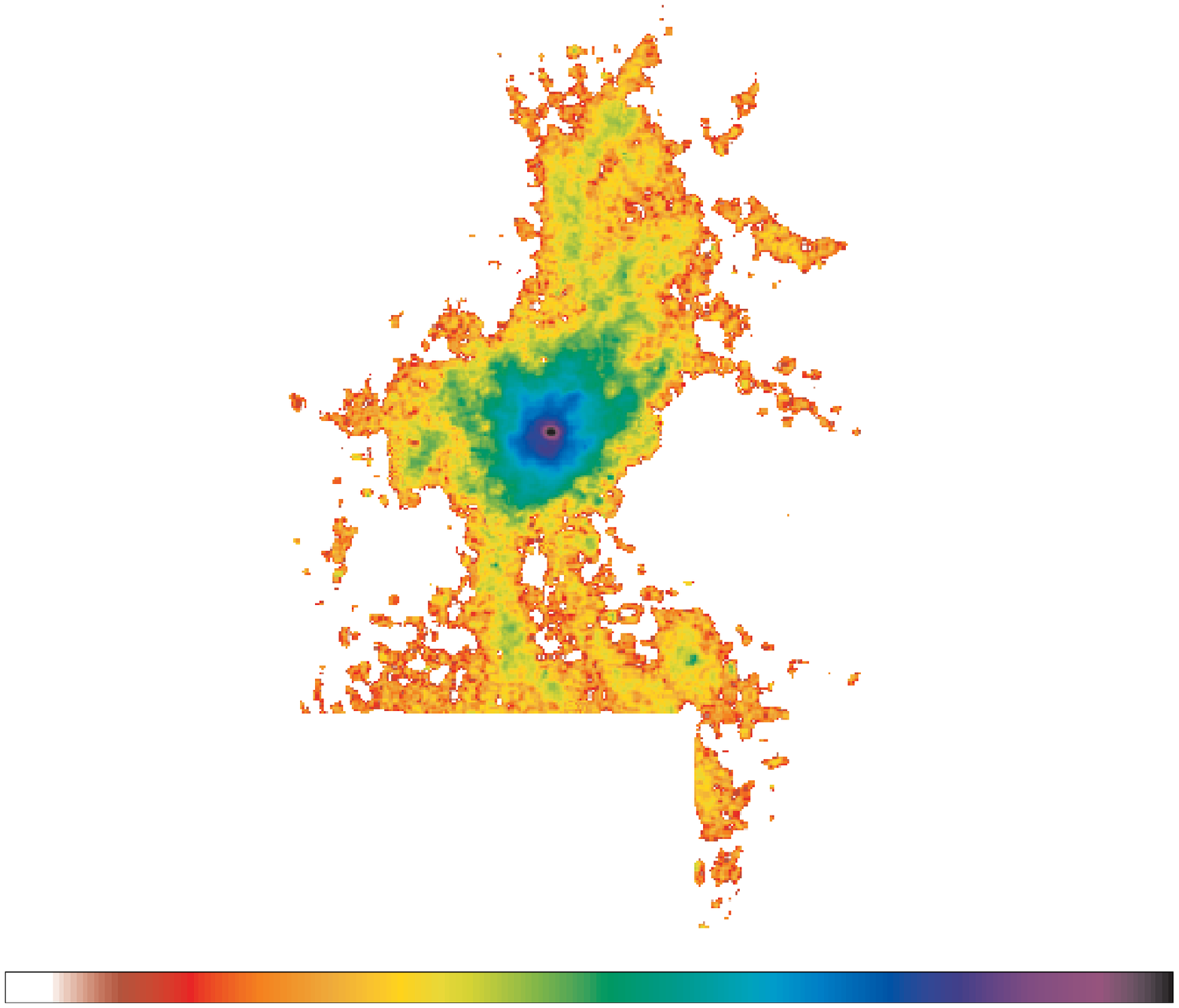,height=0.28\textheight}}
\parbox{0.5\textwidth}{
\psfig{figure=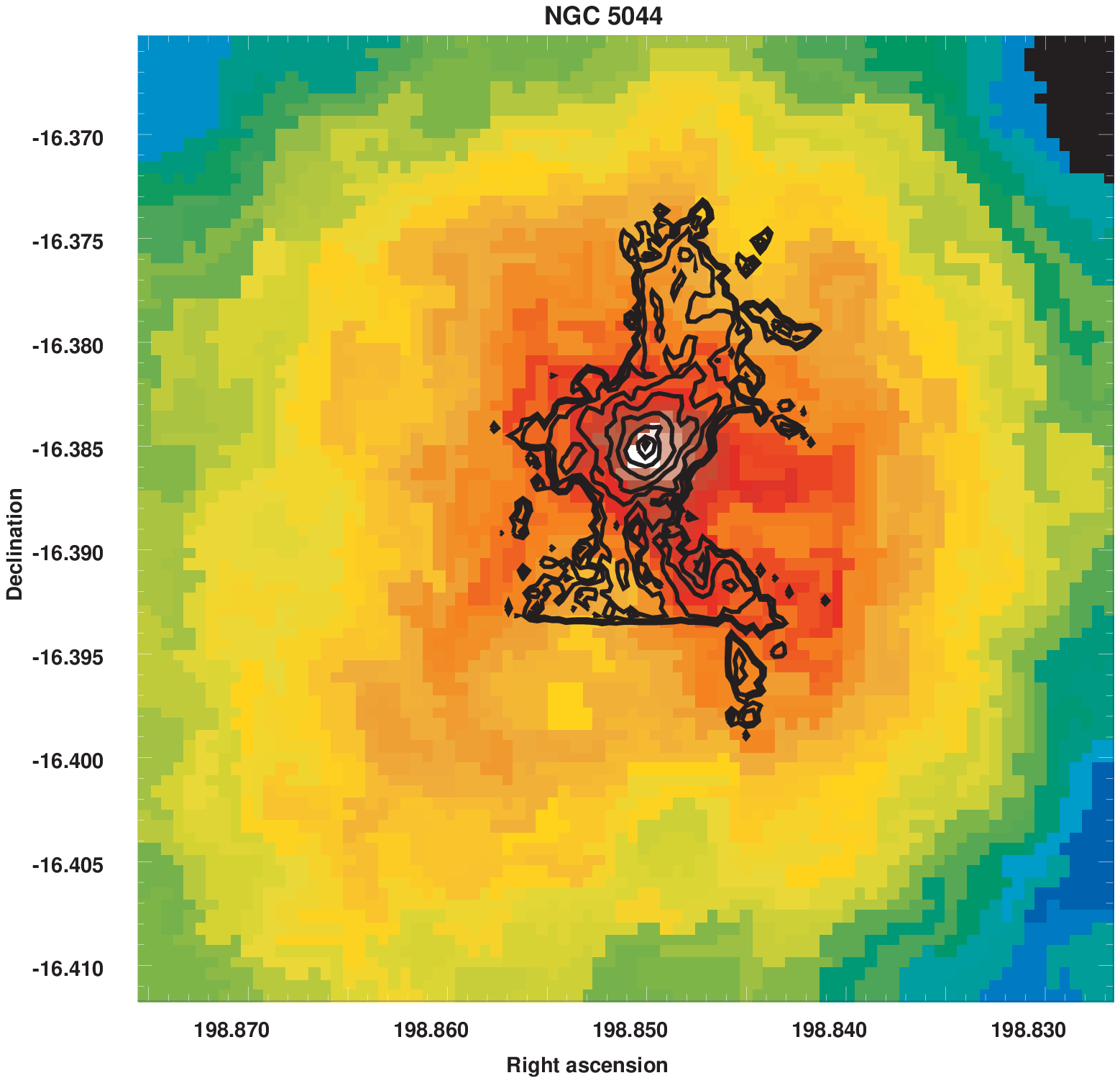,height=0.3\textheight}}
}
\caption{\label{fig.halpha} \footnotesize
\emph{Left Panel:} H$\alpha$ + [N \tiny $\rm{II}$ \footnotesize]  map of \source\ taken from 
\citet{Caon.ea:00}. \emph{Right Panel:} Surface brightness contours of the 
H$\alpha$ + [N \tiny$\rm{II}$\footnotesize] map superimposed on the contour binned image of 
Fig.\ref{fig.source}.
}
\end{figure*}

The \chandra\ data show the impact of this energy release on the hot X-ray
emitting gas which is undergoing short-term gas-dynamical activity. 
The detection of a pair of X-ray cavities allow us to give an estimate of the 
amount of energy deposited from the AGN and the time of its release, following
a now standard analysis \citep[e.g.][]{Birzan.ea:04,Dunn.ea:06}.
For each of the two cavities a size and position was measured, assuming the
projected shape is an ellipse measured from the exposure-corrected, 
un-processed image as depicted in Fig.\ref{fig.cavities}. 
We take as a guide the cavity on the south
because of the higher contrast and the surface brightness profiles of 
Fig.\ref{fig.profiles}. The two cavities are at a projected distance $R$ 
of 33\arcsec\ (6.1 kpc) from the center of the bubble to the center of 
\source, with a projected semi-major axis $a$ of 17\arcsec\ (3.1 kpc) and a 
projected semi-minor axis $b$ of 9\arcsec\ (1.7 kpc). 
We do not consider here the dip in
surface brightness at PA 212-244 because it not a clear deviation from 
the azimuthally averaged surface brightness profile. 

Assuming that the cavities are bubbles completely devoid of gas at the local 
ambient temperature and taking the pressure and temperature of the gas 
surrounding the cavities
to be the azimuthally averaged values at the projected radius of its center as
calculated in the analysis of \citet{Gastaldello.ea:07*1}, we can calculate 
the $P{\rm{d}}V$ work done on the X-ray gas by the AGN in inflating the 
cavities. The actual third axis $c$ of the ellipsoid is unknown: we assume a 
prolate shape ($b=c$). In this case we obtain 
$pV = 2.29\pm0.08 \times 10^{55}$ ergs for each cavity (the gas 
pressure at 6.1 kpc is $2.08\pm0.07 \times 10^{-11}$ dyne cm$^{-2}$).
Following \citet{Birzan.ea:04} we define three different time scales and therefore potential
ages for the bubbles: the sonic time scale $t_{c_s} = R/c_s$, where $c_s = 445\pm3$ km/s 
taking $kT = 0.77\pm0.01$ at 6.1 kpc; the buoyancy time scale $t_{buoy} = R/\sqrt{SC/2gV}$ 
where $V$ is the volume of the bubble, $S$ is the cross section of the bubble and $C=0.75$ is 
the drag coefficient \citep{Churazov.ea:01}. The gravitational acceleration $g$ can be 
calculated either as $g=GM(<R)/R^2$ using the mass profile derived in 
\citet{Gastaldello.ea:07*1} or, as there can be reasonable doubts about the strict applicability 
of hydrostatic equilibrium in the inner kpc of \source\ \citep[we are likely 
underestimating mass and therefore overestimating $t_{buoy}$, as we argued 
in][given also the failure to detect the stellar mass of the central 
galaxy]{Gastaldello.ea:07*1}, we use the stellar velocity dispersion of the
central galaxy, under the approximation of an isothermal sphere and calculate 
$g \approx 2\sigma^2/R$ \citep{Binney.ea:87}; the time required to refill the displaced volume 
as the bubble rises upward, $t_r = 2\sqrt{r/g}$, again using for $g$ either the
X-ray determined mass or the stellar dispersion velocity.
The calculated order-of-magnitude estimates for the time scales described above 
are shown in Table \ref{tab.powers}.
and they are consistent with the short
sputtering lifetime for the extended dust, $\sim 10^7$ yrs, discussed in
\citet{Temi.ea:07*1} and point towards the same episode of AGN feedback.

We can compare estimates of the energy released by the AGN to the radiative
losses within the cooling region, here estimated to be the region within
a radius where the cooling time is less than 3 Gyr \citep[e.g.][]{Dunn.ea:06},
also tabulated in Table \ref{tab.powers} which corresponds to 27 kpc. 
The task of deriving the outburst
energy from the observable $pV$ is complicated by various unknowns, like
the adiabatic index $\gamma$ of the material inside the bubble 
\citep[$\gamma=5/3$ if thermal or $\gamma=4/3$ if non-thermal, see for example the
discussion in][]{Mathews.ea:08} or the correct hydrodynamic evolution of the
bubble expansion, like its initial over-pressure compared to the ICM or the
possibility of continuous inflation, to explain the paradoxical result of
bubble energy increasing with distance from the center of the gravitational
potential \citep{Diehl.ea:08}. Using the commonly adopted recipe of estimating
as $4pV$, with $\gamma=4/3$, the energy input arising from the cavities alone,
this energy can be at most $\sim8\times10^{41}$ \ergsec, which falls short to
counteract cooling losses.

It is remarkable that a pair of cavities close to the nuclear source are 
lacking extended high frequency radio emission. This is contrary to what is 
generally observed in particular in clusters of galaxies. 
Analysis of VLA observations
at 1.56 and 4.9 GHz confirms the presence of just a point source with 
beam-deconvolved size $< 3$\arcsec and with a very flat ($\alpha \sim 0$) 
spectral index (Giacintucci et al., in preparation). 
Two other sources sharing the same behavior are 
NGC 4636 \citep[e.g.,][]{JonesC.ea:02} and NGC 4552 \citep{Machacek.ea:06}.
However the cavities in these sources have been interpreted as caused by a 
nuclear outburst which is directly causing shock heating, also because the
brighter rims surrounding the cavities are actually hotter than the 
surrounding medium, rather than displacement by the radio lobes 
\citep{JonesC.ea:02}. Or the cavities could just be related to a previous 
outburst and might just be ghost cavities, as 
suggested by \citet{OSullivan.ea:05*1} for NGC 4636 and the same reasoning
could apply to NGC 5044. In that case it remains to be explained why the
synchrotron emission has faded to undetectable values in these sources and
not in cavities with approximately the same ages in many other sources 
\citep[e.g.,][]{Birzan.ea:04}. Other ghost cavity systems with a weak central radio source
bearing no obvious relation to the observable cavities which are close to the central
galaxy are HCG 62 \citep{Morita.ea:06} and NGC 741 \citep[][where only one cavity is present]{Jetha.ea:08}. 
Sensitive low-frequency radio observations will shed further light on this issue.

The association of dust, H$\alpha$ and soft X-ray emission showed by
\citet{Temi.ea:07*1} is also strengthened by the analysis in this paper.
The presence of the N cavity and in particular of the relatively cooler 
emission in the NW filament (see \S\ref{subsection_spectral.core})
likely explains the origin of the N H$\alpha$ filament. X-ray filaments are present at 
both sides of the cavities, but only
the ones with the presence of dust are showing optical emission and 
\emph{cooler} X-ray emission (see Fig.\ref{fig.halpha} and the analysis of 
\S\ref{subsection_spectral.core}). \citet{Temi.ea:07*1} showed that the
cospatiality of these features can be explained as the result of dust-assisted cooling
in an outflowing plume of hot dusty gas: dust can cool buoyant gas to $10^4$ K, which
emits the optical emission lines observed. The warm-gas phase is maintained in
thermal equilibrium near $\sim10^4$ K by radiative losses and likely a combination
of thermal conduction and UV heating from post-AGB stars 
\citep[see the calculation of Tab.5 in][ ionization from post-AGB stars 
could explain only 42\% of the optical line emission in \source]{Macchetto.ea:96}.
It is unfortunate that the deeper H$\alpha$ observation of \citet{Caon.ea:00}, compared to the
one by \citet{Goudfrooij.ea:94} used in the comparison of \citet{Temi.ea:07*1},
is affected by a CCD defect in the southern region co-spatial to the X-ray 
cavity: it looks like the nebular emission is more extended and can cover
also the cavity.
Another tantalizing evidence of the association
of these three components comes for example from NGC 5846 
\citep[][and references therein]{Trinchieri.ea:02}. Evidence of PAH emission in
\spitzer\ IRS spectra has been found also in NGC 1275 \citep{Johnstone.ea:07}.
It has been discussed that AGN feedback can provide
some heating (negative feedback), but it can also be responsible for 
a positive feedback, i.e. for inducing cold gas production 
\citep[e.g.,][]{Pizzolato.ea:05,Revaz.ea:08}. In this scenario AGN feedback
itself is responsible for the production of overdense blobs which cool
rapidly producing H$\alpha$ emitting gas, molecular gas and star formation.
Dust transported from the dusty disks of the central elliptical into
the ICM by the episode of AGN feedback acts as a catalyst for the cooling 
of the gas.
It would be therefore interesting for the recent renewed theoretical 
interest in the generation and survival of the optical line emission filaments 
\citep{Nipoti.ea:04,Revaz.ea:08,Pope.ea:08} to include the neglected 
dust-assisted cooling in the energy balance equation. 

\subsection{The cold fronts and the dynamical state of the group NGC 5044}

The spectral analysis presented in  \S\ref{subsection_spectral.coldfronts} 
indicated that the pair of surface brightness discontinuities detected
in the \chandra\ and \xmm\ images are cold fronts.
Given the large scale relaxed morphology of
\source\ and its rising entropy profile \citep{Gastaldello.ea:07*2}
as commonly observed for bright relaxed groups, the source nicely fits
in the explanation for the emergence of these features proposed by 
\citet{Ascasibar.ea:06} as due to gas sloshing caused by an off-axis merger with a 
smaller satellite.
Another tell-tale sign of this encounter, as also suggested by 
\citet{Ascasibar.ea:06}, is the presence of a peculiar velocity of the
central galaxy, which is often exhibited by central cD galaxies in relaxed 
clusters \citep[e.g.,][]{Oegerle.ea:01}. 
Indeed NGC 5044 is known to have a $\sim$ 150 km/s peculiar velocity 
with respect to the mean group velocity 
\citep[][and references therein]{Mendel.ea:08}. A sub-clump of galaxies statistically 
significant to the Dressler-Schectman test \citep{Dressler.ea:88}  
has been detected in the north-east outskirts of the \source\ \citep{Mendel.ea:08}
which might be the smaller merging sub-group responsible for the gas sloshing and 
the cold fronts. For a virial mass for \source\ of $3.7 \times 10^{13}$ \msun\ 
\citep{Gastaldello.ea:07*1}, assuming a mass ratio of 5 as in the reference
case of \citet{Ascasibar.ea:06} 
would imply a mass of $\sim 8 \times 10^{12}$ \msun\ for the satellite.
If this candidate sub-clump is real, it would corroborate the proposed scenario and the
mechanism proposed by \citet{Ascasibar.ea:06} for the formation of
cold fronts \citep[see also the case of the cluster A 496,][]{Dupke.ea:07} 

Surface brightness discontinuities as cold fronts have so far been 
discussed in many merging elliptical galaxies and groups 
\citep[e.g., NGC 1404,][]{Machacek.ea:05}. An unusual discontinuity 
for the group NGC 507, interpreted as an 
abundance jump, closely related to the expansion of a radio lobe, has been discussed
by \citet{Kraft.ea:04}. NGC 5044 is to our knowledge the first relaxed group for which
cold fronts have been discussed in close similarity to the ubi\-qui\-tous ones 
detected in relaxed clusters. Further examples, given enough data quality, are likely
to be discovered: suggestions have already been made for objects like MKW 4 and
IC 1860 \citep{Gastaldello.ea:07*1}.
%

\section{Conclusions}
\label{conclusion}

We have presented results of a two-dimensional analysis of the currently available
\chandra\ and \xmm\ data for the bright nearby galaxy group \source. The results
can be summarized as follows:

\begin{itemize}
\item A pair of X-ray cavities have been detected, further confirming the recent
outburst indicated by the extent and morphology of H$\alpha$ and dust emission.
\item The presence of cooler filamentary X-ray emission co-spatial with H$\alpha$ and dust emission strengthens our previous suggestion that dust-aided cooling could 
contribute to the H$\alpha$ emission.
\item The presence of a set of two cold fronts together with a peculiar velocity
of the central galaxy NGC 5044 suggests a disturbance of an overall relaxed system
by an off-axis merger with a smaller satellite. 
\end{itemize} 

The detection of such a rich phenomenology has been possible due to \source\
being bright and nearby, much like on bigger mass scales has been possible
with M87 and Perseus. As for these objects up-coming deeper X-ray observations 
with \chandra\ \citep{David:07} and \xmm\ \citep{Kaastra:07} will shed 
further light on the properties of \source\ and groups of galaxies in general. 
Multi-wavelength observations, i.e. in H$\alpha$ and radio bands, are needed 
for a deeper understanding of the physical processes in the core.

\begin{acknowledgements}
We would like to thank N. Caon for kindly providing the H$\alpha$ image used 
in Figure \ref{fig.halpha} and for useful discussions; S. Giacintucci for 
sharing her results prior to publication and for useful 
discussions; P.J. Humphrey for the use of his \chandra\ data reduction and 
analysis code; J. Sanders for the use of his contour binning code; 
L. Zappacosta for a critical reading of the manuscript; the anonymous referee
for a carefulr reading and for suggestions that improved the work
presented here.
We made use of the WVT binning algorithm by Diehl \& Statler (2006), which is 
a generalization of Cappellari \& Copin's (2003) Voronoi binning algorithm.
F.G., F.B. and S.E. acknowledge the financial contribution from contract 
ASI-INAF I/023/05/0 and I/088/06/0. D.A.B. gratefully acknowledges partial support 
from NASA grant NNG04GE76G, issued through the Office of Space Sciences 
Long-Term Space Astrophysics Program.
\end{acknowledgements}

{\it Facilities:} \facility{CXO}, \facility{XMM}
%
%
%
\bibliographystyle{apj}
\bibliography{gasta}

\end{document}